\documentclass[%
 reprint,
nofootinbib,
 amsmath,amssymb,
 aps,
]{revtex4-1}

\usepackage{graphicx}
\usepackage{dcolumn}
\usepackage{bm}

\begin{document}

\preprint{APS/123-QED}

\title{ A violation of the covariant entropy bound?}

\author{Ali Masoumi}
 \email{ali@cosmos.phy.tufts.edu}
\affiliation{Tufts Institute of Cosmology,
255 Robinson Hall, 212 College Ave,
Medford, MA, 02155
}%

\author{Samir D. Mathur}
 \email{mathur.16@osu.edu}
\affiliation{Department of Physics, The Ohio State University, Columbus,
OH 43210, USA
}%

\def\({\left(}
\def\){\right)}
\def\[{\left[}
\def\]{\right]}

\def\nn{\nonumber \\}
\def\p{\partial}
\def\t{\tilde}
\def\h{{1\over 2}}
\def\be{\begin{equation}}
\def\bea{\begin{eqnarray}}
\def\ee{\end{equation}}
\def\eea{\end{eqnarray}}
\def\b{\bigskip}


\begin{abstract}

Several arguments suggest that the entropy density at high energy density $\rho$  should be given by the expression $s=K\sqrt{\rho/G}$, where $K$ is a constant of order unity. On the other hand the covariant entropy bound requires that the entropy on a light sheet be bounded by $A/4G$, where $A$ is the area of the boundary of the sheet. We find that in a suitably chosen cosmological geometry,   the above expression for $s$ violates the covariant entropy bound. We consider different possible explanations for this fact; in particular the possibility that entropy bounds should be defined in terms of volumes of regions rather than areas of surfaces.

\end{abstract}

\keywords{Black holes, string theory}
\maketitle


\section{\label{secone}Introduction}

How much entropy $S$ can there be in a given region? The `Bekenstein bound'  \cite{bek2} was proposed for regions where gravity is not strong. It says that $S< 2\pi ER$, where $E$ is the energy in the region and $R$ is some measure of its physical size. The Bekenstein-Hawking entropy  of black holes is given by $S={A/ 4G}$ \cite{bek}; based on this expression it has been argued that in strongly gravitating systems the number of degrees of freedom is proportional to the surface area of a region rather than the volume. 

Cosmology poses a further challenge.  Consider a flat universe at an early time in its evolution; say, in the radiation dominated phase. Consider a box-shaped region of physical volume $V$. the entropy in the box is proportional to $V$, and so for a sufficiently large box the entropy $S$ in the box will exceed the surface area of the box. Are there any limits that we can place on $S$?

\begin{figure}[h]
\includegraphics[scale=.32]{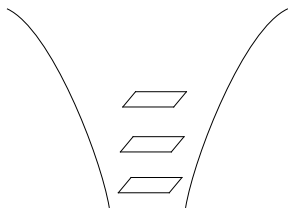}
\caption{\label{fone} A box of the same physical size at different times in an expanding cosmology. At an early enough time, the box  will contain more mass than required to make a black hole with size equal to the size of the box. }
\end{figure}

One can hope to bound the entropy in the box by limiting the box itself. Bekenstein \cite{bek3} argued that entropy bounds should be plced on  the matter inside the particle horizon. In \cite{fs}, it was argued that entropy bound should be given by $A/4G$, were $A$ is the area of the particle horizon.  In \cite{venezianopre,veneziano2}, it was argued that the more correct scale for this purpose is $H^{-1}$, where $H$ is the Hubble constant.   In \cite{br} it was suggested that  one should use the scale set by the apparent horizon arising from the cosmological expansion\footnote{In \cite{el} it was argued that entropy bounds of this form should not be considered; instead the most general constraint on entropy comes from the generalized second law of thermodynamics. In \cite{rs} it was noted that the proposal of \cite{fs} gives constraints of the fluctuations arising from inflation.}; this scale is also $\sim H^{-1}$.
 In \cite{bousso} a covariant version of the entropy bound was given: the entropy passing through a `light sheet' should be less than $A/4G$, where $A$ is the area of a surface that bounds the light sheet.  

In this paper we will consider a certain expression for the entropy density $s$, which is motivated by several different physical considerations. We will then observe that this $s$ violates the apparent horizon bound and the covariant entropy bound. 

In more detail, we proceed as follows:

\b

(i)  We recall that several different considerations lead to the expression
\be
s=K\sqrt{\rho\over G}
\label{density}
\ee
for the entropy density in the early universe. Here $\rho$ is the energy density, and $K$ is a constant of order unity, determined by the details of the gravitational theory. In string theory, it appears plausible that one can actually construct a set of states that give rise to such an $s$; we will recall this construction. For example we can take a lattice of string states \cite{veneziano} that are near the Horowitz-Polchinski correspondence point. 

\b

(ii) We consider a flat cosmology
\be
ds^2=-dt^2+\sum_{i=1}^d a_i^2(t) dx_i dx_i
\ee
At some time $t_0$ we choose initial conditions 
\be
{\dot a_i(t_0)\over a_i(t_0)}\equiv b_{i0}
\label{intro1}
\ee
to get an expansion that is in general asymmetric, since the $b_{i0}$ need not be all equal. We will need to consider only an infinitesimally thin  slice of this cosmology
\be
t_0-\Delta t ~<~t~<~t_0, ~~~~~~\Delta t \rightarrow 0
\label{thinslice}
\ee
The metric of this thin slice is determined just by the choice of initial conditions (\ref{intro1}). The only constraint on the $b_{i0}$ comes from the Friedmann equation relating them to $\rho$. 

\b

(iii) We construct the a light sheet in the thin slice (\ref{thinslice}), and compute the entropy flowing through this light sheet assuming the entropy density is given by (\ref{density}). We find that the covariant entropy bound of \cite{bousso} is violated if the expansion is suitably asymmetric. For example, suppose all but one of the $b_{i0}$ are equal: $b_{i0}=b, i=2, \dots, d$, while $b_{10}$ is allowed to be different. Then the bound is violated if
\be
{b_{10}\over b} ~>~ {{\pi \over 2 K^2 }(d-1) }  - {d-2 \over 2}~.
\label{wwone}
\ee
This is the main result of this paper. To summarize, it appears that we can construct states in string theory to get the entropy density (\ref{density}), and this entropy density violates the covariant entropy bound in the above mentioned asymmetric cosmological metric. 

\b

Note that we are not trying to violate the bound by assuming that $K$ is larger than a certain value. Rather, we are assuming that the theory of gravity determines some value of $K$, and then we show that for {\it any} such $K$, the bounds will be violated when the initial conditions of the expansion are chosen to be sufficiently asymmetric. Note that the asymmetry required is not parametrically large in any way; it is just order unity. For example, if $d=3$ and $K=1$, then the bound is violated for
\be
{b_{10}\over b}>2.65
\ee

After arriving at  (\ref{wwone}),  we explore other issues related to this result. In particular we look at entropy bounds based on the apparent horizon, and on causal connection scales; we find that the former is violated by asymmetric expansion while the latter need not be.

What should we conclude from this violation of the bounds? There are several possibilities:

(i) It might be that some physical  effect in string theory disallows the states leading to (\ref{density}), or disallows the asymmetric initial conditions that we assume. In that case we would learn something interesting about the limitations of the expression (\ref{density}). 

(ii) It might be that the conditions assumed in the entropy bounds need to be tightened, so that the states leading to (\ref{density}) are not appropriate candidates for application of the bound. In that case we  might discover new limitations on the application of entropy  bounds to the very early universe. 

(iii) It might be that the very notion of bounding entropy by area is flawed; in that case it may be that (\ref{density}) itself gives the general bound on entropy.

 \bigskip
 
The plan of this paper is as follows. In section\,\ref{sectwo} we recall the arguments for the equation of state (\ref{density}). In section\,\ref{secbound} we review the covariant entropy bound, using as an illustration the geometry which we will use later when examining the bound. 
Section\,\ref{infinitesimal} contains the main result of this paper: we show that the covariant entropy bound is violated in a thin slice geometry for the equation of state (\ref{density}) if we consider suitably anisotropic expansion. In section\,\ref{full} we compute the full evolution for the equation of state $p=\rho$, and again note the violation of the covariant entropy bound. In section\,\ref{viscous} we discuss the evolution in the situation where we have $p=\rho$, but also the minimum viscosity required by the conjectured bounds on $\eta/s$. In section\,\ref{apparent} we perform a check for the entropy bound based on the apparent horizon, and again find a violation for suitably asymmetric expansion. In section\,\ref{marolf} we recall  some proofs that were proposed for  the covariant entropy bound. These proofs assumed  certain conditions on the entropy flow, but we find that the  asymmetric initial conditions we need for violating the bound also violate the assumptions made in these proofs; thus our results are not in conflict with these proofs. In section\,\ref{ceb} we recall the causal entropy bound proposed in \cite{brusv} and note that it can be consistent with the equation of state (\ref{density}). Section\,\ref{discussion} is a discussion where we analyze various possible implications of our results.

\section{A conjecture for the entropy at high densities}\label{sectwo}

In this section we will review the arguments that lead to a certain suggestion for the entropy density $s$ of the early universe. We will first explain the construction of the states in string theory that lead to this expression for $s$. We will  then note that several abstract lines of argument (not necessarily related to string theory) suggest the same expression for $s$. 

\subsection{The entropy in a box}\label{onea}

Consider a toroidal box of volume $V$. In this box we put an energy $E$. What is the entropy 
\be
S=S(E,V)
\ee
in the limit where the energy density $\rho=E/V$ becomes large?

\begin{figure}[h]
\includegraphics[scale=.72]{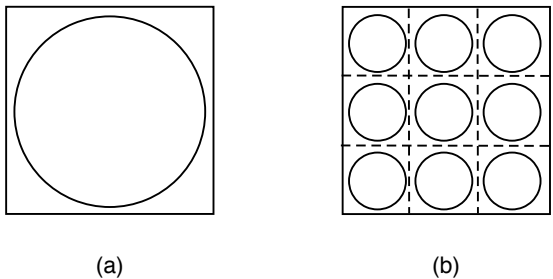}
\caption{\label{fbh} (a) At energy $E\sim E_{\rm bh}$ the maximum entropy is attained by a single hole filling the box (b) At larger $E$, a lattice of black holes has more entropy than a single hole.}
\end{figure}

 For low values of the  $E$,   we  expect the phase of matter to be radiation. This phase has  entropy $S\sim V\rho^{d/ (d+1)}$, where $d$ is the number of spatial  dimensions. At larger $E$, we can get more entropy by forming a black hole. As we increase $E$, we reach a critical value 
$E\sim E_{\rm bh}$, where the radius of the hole $R_s$ becomes order the size $L$ of the box (fig.\ref{fbh}(a)). The entropy of this black hole is 
\be
S_{\rm bh}= {A\over 4G}\sim {L^{d-1}\over G} 
\ee
Can we fit more entropy than $S_{\rm bh}$ in our box with volume $V\sim L^d$? A simple construction shows how we can achieve such a goal. Instead of one black hole filling $V$, we just consider a lattice of smaller holes (fig.\ref{fbh}(b)). We will recall below the computation of entropy for such a lattice. But before we do that, it is helpful to address a few natural questions: 

\bigskip

(a) It is often said that entropy is maximized when all the energy is put into a single black hole. But this statement is true only if we hold the total energy fixed, and allow the hole to expand to its natural size in an asymptotically flat space.  Suppose instead that we fix the volume to a given value $V$, and do {\it not} constrain the energy $E$. Then we can get more entropy than that of a single large hole, as can be seen from fig.\ref{fbh}. In fig.\ref{fbh}(a), we have a single large hole filling the box; the surface area of this hole is of order the area $A_{\rm box}$ of the walls of the box. In fig.\ref{fbh}(b) we have a lattice of smaller holes. The total area of these holes is of order the total surface area of the cubes that contain these holes. The total surface area of these cubes includes a contribution $A_{\rm box}$ from the outer walls of the box. But we also get the contribution of all the {\it other} walls depicted by the dashed lines, so the total surface area of these cubes $A_{\rm cubes}$ is much more than $A_{\rm box}$. Thus we see that the entropy of the lattice of holes in fig.\ref{fbh}(b) can be made much more than the entropy of the single hole in fig.\ref{fbh}(a).

\bigskip

(b) Even though the lattice of black holes in fig.\ref{fbh}(b) has more entropy than the single black hole of fig.\ref{fbh}(a), one might wonder if the lattice of holes is somehow unstable to collapsing into a single large hole. But this is clearly impossible, since the lattice of holes has much more entropy than the single hole. The holes can certainly move and interact, but the configuration they will end up in cannot be a single black hole; the total entropy is $\sim A_{\rm cubes}/G$ which is much more than $A_{\rm box}/G$.

\bigskip

 \begin{figure}[h]
\includegraphics[scale=.42]{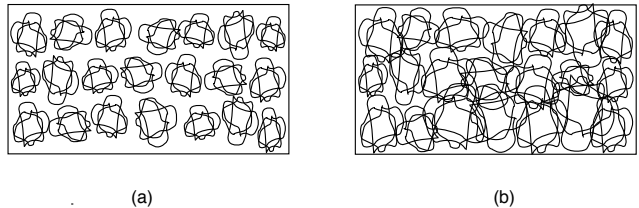}
\caption{\label{ffour2} A pictorial depiction of the configurations that reproduce the entropy (\ref{density}). (a) Clusters of intersecting branes give the entropy of order the black hole entropy for each cluster. The overall entropy is then the sum of these entropies. (b) These brane sets will in general interact and merge, but the entropy cannot decrease below the value obtained from the separate brane sets of (a).}
\end{figure}

(c) Black holes might appear to be somewhat esoteric objects, and one might wonder if entropy bounds are required to work for a fluid made of such holes. But in string theory we know that the entropy of a black hole can be reproduced by considering  a set of intersecting branes \cite{sv,hms}; further, the size of this set is of order the horizon radius of the corresponding black hole \cite{fuzzballsize}. In fig.\ref{ffour2} we depict a lattice made of sets of such intersecting branes. It is true that the intersecting brane construction in string theory is best established for extremal and near extremal holes, but this is sufficient for our purpose. We can let the lattice in fig.\ref{ffour2}(a) be composed of extremal holes, with charge alternating between positive and negative for successive holes in the lattice.  Then the overall configuration will be charge neutral as required for a cosmology.  The resulting configuration of branes  is a configuration of `normal matter' in string theory, and gives an entropy of the same order as the entropy $S_{\rm cubes}$ in fig.\ref{fbh}(b). It is true that if we start with the configuration if fig.\ref{ffour2}(a) and then let it evolve, the  brane sets  will in general interact and merge to create  more complicated states (fig.\ref{ffour2}(b)). But since entropy can only increase, the resulting configurations will not have an entropy that is below the entropy of fig.\ref{ffour2}(a), which is of order $S_{\rm cubes}$. 

\bigskip

\subsection{The entropy of a lattice of black holes}\label{lattice}

Let us now compute the entropy of the lattice of black holes in fig.\ref{fbh}(b).

We will assume that we are in a $(d+1)$-dimensional  spacetime. Consider a  torus $T^d$ with volume $V$. Consider a lattice of black holes, each with the same radius $R_s$. Let the  separation between the holes also be $\sim R_s$. The number of holes is then
\be
N_{\rm hole}\sim \left ( {V\over R_s^d}\right )
\ee
The entropy of each hole is
\be
S_{\rm hole}\sim {R_s^{d-1}\over G}
\ee
Thus the total entropy is
\be
S\sim N_{\rm hole} S_{\rm hole}\sim {V\over R_s G}
\label{entropylattice}
\ee
We see that we can make $S$ as big as we want by making $R_s$ small enough. In particular,  the entropy of such configurations can exceed the entropy given by the surface area of the box. The energy of each hole is
\be
E_{\rm hole}\sim {R_s^{d-2}\over G}
\label{eseven}
\ee
Thus the total energy is
\be
E\sim N_{\rm hole} E_{\rm hole}\sim {V\over R_s^2 G}
\label{elattice}
\ee
From this expression we have
\be
R_s\sim \left ( {V\over EG}\right )^\h
\ee
Substituting this in (\ref{entropylattice}) we find
\be
S\sim {1\over R_s} {V\over G}\sim \left ( {V\over EG}\right )^{-\h}{V\over G} \sim \sqrt{EV\over G}
\ee
Introducing a constant $K$ of order unity we get
\be \label{eq:EntropyFormul}
S=K\sqrt{EV\over G}
\ee
Noting that
\be
\rho={E\over V}
\ee
we see that
\be
S=K\sqrt{\rho\over G} \, V
\label{oneqq}
\ee
The entropy density $s=S/V$  is then given by (\ref{density}). 

\subsection{General arguments for the entropy expression (\ref{density})}\label{argument}

Let us now recall the various arguments that suggest the expression (\ref{density}) (or equivalently, eq.(\ref{oneqq})) for the entropy density of the early universe.

\b

(a) In the above discussion we treated our box  as having a fixed size. But in fact the energy density $\rho$ will cause the box to expand in accordance with the Friedmann equation. This expansion leads to a cosmological horizon, with radius
\be
 H^{-1}\sim (G\rho)^{-\h}
\ee
Supose we argue that   the entropy in a region of radius  $H^{-1}$ should be given by the entropy of a black hole with radius $\sim H^{-1}$;  models of this kind can be found in \cite{veneziano2,bf, veneziano}.    The entropy of a black hole of radius $ H^{-1}$ is $S\sim H^{-(d-1)}/G$. If we use a similar expression for the entropy in a cosmological horizon region, then the entropy density would be
\be
s\sim {S\over H^{-d}}\sim {H\over G }\sim \sqrt{\rho\over G}
\label{fsbound}
\ee
in agreement with (\ref{density}). 

\bigskip

(b) Banks and Fischler made a detailed study of a universe where the energy is in the form of a dense collection of black holes - a `black hole gas' \cite{bf}.  It was noted that the entropy (\ref{density}) corresponds to an equation of state 
\be
p=\rho
\label{eos2}
\ee
 A general picture was developed where horizon sized black holes coalesce as the universe expands, so that the entropy in a  region of size $H(t)^{-1}$ remains of order the entropy of a  black hole of radius $H(t)^{-1}$. 
 
 When $\rho$ is of order the string scale, it was argued in \cite{veneziano} that  an entropy density (\ref{density}) would be obtained for a closely packed gas of string states which are at the `Horowitz-Polchinski correspondence point' \cite{hp} (i.e., at the point where the string is large enough to be at the threshold of collapsing into a black hole).
 
 \bigskip
 
 (c) The expression (\ref{density}) was obtained in \cite{sas1} by arguing for a `spacetime uncertainty relation'.   In \cite{brusv, brus2} the notion of a causal connection scale was used to arrive at the same equation of state (\ref{density}). In \cite{verlinde} a similar relation was argued to correspond to the Cardy formula for the density of states.

 \bigskip
 
 (d) In \cite{sas2} it was noted that the expression (\ref{oneqq}) was invariant under the T and S dualities. It was noted that if one further requires that $S\propto V$ (i.e. $S$ is extensive), then we get (\ref{oneqq}). 
 
 In \cite{alimathur} the expression (\ref{oneqq}) was obtained as an extension of the entropy of black holes. Let $E_{\rm bh}$ be the energy for which a black hole has  a size of the order of our torus. Suppose we require that
  
(i) $S$ should be invariant under T-duality in any cycle of the torus.
 
  (ii) $S$ should be invariant under S-duality.
  
   (iii) We should get $S\sim S_{\rm bh}$ when the box size and shape is such that $E\sim E_{\rm bh}$ for that box.

Then it was argued in \cite{alimathur} that we are led to the expression (\ref{oneqq}),
in the domain
\be
\rho_{\rm bh}\lesssim \rho\lesssim \rho_{\rm p}~,
\label{two}
\ee
Here $\rho_{\rm bh}$ is the energy density corresponding to the energy $E_{\rm bh}$ placed in the volume $V$, and $\rho_{\rm p}$ stands for the Planck density -- Planck mass per unit Planck volume. At the lower end of this range ($\rho\sim \rho_{\rm bh}$) the expression (\ref{oneqq}) matches onto the area entropy of the black hole $S_{\rm bh}\sim A/4G$. At the upper end $\rho=\rho_{\rm p}$, (\ref{oneqq}) gives an entropy of one bit per unit Planck volume.\footnote{We will recall the details of this statement in section\,\ref{discussionc} below.} Thus (\ref{oneqq}) extrapolates the Bekenstein `area entropy'   to the domain (\ref{two}). Since $\rho \gtrsim \rho_{\rm bh}$, we will say that matter is `hyper-compressed'; i.e., compressed beyond the density of the largest black hole that can fit in the box.

\bigskip

It is interesting that the different approaches (a)-(d) mentioned above all lead to the expression (\ref{density}).\footnote{For other conjectures about  the entropy in the early universe, see for example \cite{bv,branegases,rama}. For other conjectires on the entropy in gravitaional systems, see \cite{marolf}.}

 \subsection{The equation of state}

Let us compute the values of different thermodynamical quantities that follow from the equation of state
  \be
  S=K\sqrt{EV\over G}
  \label{oneqq2}
  \ee
   The observations below were noted earlier in \cite{sas1,bf}, and a detailed dynamics was conjectured for such an equation of state in \cite{bfm}.  

  The first law of thermodynamics gives
  \be
  TdS=dE+pdV
  \ee
  Thus
  \be
  T=\left ( {\p S\over \p E}\right ) _V^{-1}={2\over K} \sqrt{EG\over V}
  \ee
  \be
  p=T\left ( {\p S\over \p V }\right ) _E = {E\over V}=\rho
  \label{eos}
  \ee
  Writing $p=w \rho$ we see that 
  \be
  w=1
  \label{weo}
  \ee

\section{The covariant entropy bound}\label{secbound}

The expression $S=A/4G$ for the entropy of a black hole suggests that the entropy in a region is somehow limited by its surface area in Planck units. But consider a flat homogenous cosmology. Since the spatial slices are homogenous, the entropy $S$ should be proportional to the volume  of any region ${\cal R}$  on this slice. But for a large enough ${\cal R}$, the entropy in ${\cal R}$ would exceed the quantity $A/4G$ defined through the surface area of the region ${\cal R}$. 

\begin{figure}[h]
\includegraphics[scale=.52]{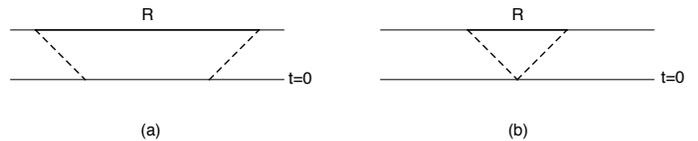}
\caption{\label{fcosmo1} The entropy in a region ${\cal R}$ increases with the volume of ${\cal R}$. To get a bound proportional to the area of the boundary of ${\cal R}$, we consider only the entropy that passes through the light sheet shown with the dotted lines. (b) If we take the region ${\cal R}$ to be a horizon volume (and assume that entropy is conserved in each comoving volume), then  the entropy on ${\cal R}$ is the same as the entropy passing through the light sheet bounding ${\cal R}$.}
\end{figure}

To remedy this problem, the following idea was suggested in  \cite{fs}.  Consider a light-like surface formed by light rays heading to the past and towards the center of ${\cal R}$. Instead of looking at the entropy in the region ${\cal R}$, we should only look at the entropy that passes through this light-like surface. It was then conjectured that this entropy would always be bounded by $A/4G$, where $A$ is the area of the boundary of ${\cal R}$. 

A somewhat different approach was suggested in \cite{venezianopre}, where it was argued that the Hubble radius $H^{-1}$ should be taken as  a `maximal box size'; the entropy in a ball of radius $H^{-1}$ should be bounded by $A/4G$ where $A$ is the area of the boundary of this ball.  In \cite{br} a suggestion was made that also involves the scale $H^{-1}$. We consider a space-like slice, with a particular point on this slice. Assuming spherical symmetry around this point, we find the radius $X_a$ of the apparent horizon. The conjecture then says that the entropy within $X_a$  is less than $A/4G$, where $A$ is the area of the apparent horizon. 

While such prescriptions were interesting, it was found that there are counterexamples to such conjectures (see for example \cite{kl}). An approach without these problems was developed by Bousso \cite{bousso}. Let us describe this proposal, called the `covariant entropy bound' or the `Bousso bound', in more detail:

\bigskip

(a) We assume that we are in $(d+1)$-dimensional spacetime. Consider a $(d-1)$-dimensional space-like hypersurface ${\cal S}$. This hypersurface may be closed (i.e. without boundary) or it may be open (i.e. with boundary). Let $A$ be the area of ${\cal S}$. 

As an illustration let us take an open hypersurface ${\cal S}$; later on we will use this example (and the notation given below) to carry out  our investigation of entropy bounds for the equation of state (\ref{density}). We let the ${\cal S}$  be a cuboid in the directions $x^2, \dots x^d$, spanning the coordinate ranges
\be
0\le x^i\le L_i
\label{hyper}
\ee
All points on this cuboid are at a fixed value of time $t$ and space coordinate $x^1$:
\be
t=t_0, ~~~x^1=x^1_0
\label{times}
\ee

\begin{figure}[h]
\includegraphics[scale=.82]{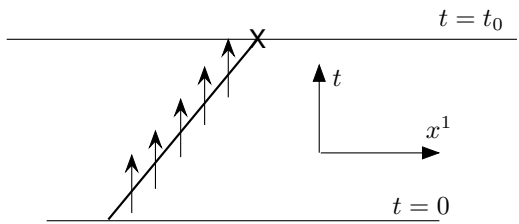}
\caption{\label{fsheet} The set-up of the covariant entropy bound in $(2+1)$-dimensional spacetime $\{ t, x^1, x^2\} $.The cross depicts the spatial direction $x^2$ which goes into the plane of the paper. The solid slanted line starting at the cross  is a null ray headed to the past; this ray stays at constant $x^2$. The set of such null rays (for different $x^2$) form the light sheet. The upward arrows depict entropy crossing the light sheet.  }
\end{figure}
\begin{figure}[h]
\includegraphics[scale=.37]{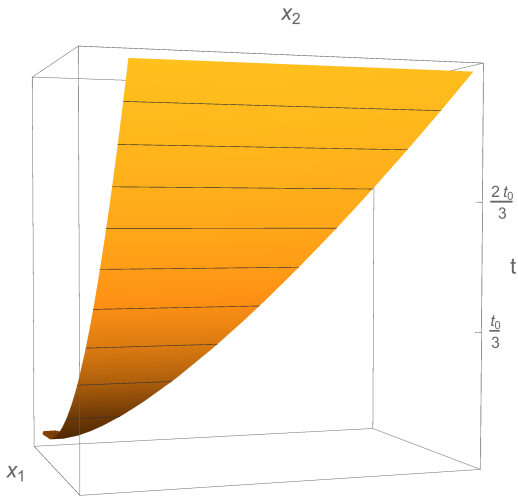}
\caption{\label{fsheet2l} The same set-up as in Fig.\ref{fsheet}, but in a 3-dimensional depiction. The length of the $x^2$ direction shrinks to zero as we follow the light sheet back towards $t=0$.  }
\end{figure}

\bigskip

(b) At each point of the hypersurface ${\cal S}$ we look for a null geodesic leaving the hypersurface, in a direction that is orthogonal to the hypersurface. In general there will be four such null geodesics from every point: there are two opposite spatial directions to move out in, and the geodesic could be future directed or past directed.   We  consider the set of null geodesics constructed this way; one from each point of ${\cal S}$. (As we will note below, there will be further restrictions on this set of null geodesics.)

In our illustration, let us assume that the metric has the form
\be
ds^2=-dt^2+\sum_{i=1}^d a_i^2(t) dx^i dx^i
\label{metricq}
\ee
so that the directions $t, x^i$ are all orthogonal to each other.  Then for the hypersurface (\ref{hyper}), the orthogonal spatial direction is $x^1$. 
In fig.\ref{fsheet} we depict the null geodesics heading in the positive direction $x^1$ as we move to the past. Each geodesic remains at a fixed value of the coordinates $x^2, \dots x^d$. The change of $x^1$ is found by requiring $ds=0$ in the metric (\ref{metricq}):
\be
{dx^1\over dt}={1\over a_1(t)}
\label{dxdt}
\ee

\bigskip

(c) We require that the set of null geodesics constructed this way be {\it nondiverging} as we move away from our hypersurface ${\cal S}$. In other words, suppose we consider a small area element $dA$ on ${\cal S}$, and the subset of the null geodesics discussed above that start in $dA$. After we follow these geodesics for an affine distance $\lambda$, the transverse area spanned by the geodesics will have a  value $dA (\lambda)$.  We require that
\be
{dA(\lambda)\over d\lambda}\le 0
\ee

In our illustration, we  can consider the area element defined by  infinitesimal intervals $dx^i,\,  i=2, \dots d$. In the metric (\ref{metricq}), the area of this element is
\be
dA(t) =\prod_{i=2}^d a_i(t) dx^i
\ee
 Requiring that $dA(t)$ decrease along the geodesics heading to the past is equivalent to
 \be
{dA\over dt}= {d\over dt} \left (\prod_{i=2}^d a_i(t)\right ) \ge 0
 \label{convergence}
 \ee
 
\bigskip

(d) We follow the null geodesics described above up to the point where they reach a caustic; i.e., the point where the separation of neighboring geodesics goes to zero. The surface spanned by the null rays emanating from ${\cal S}$, followed up to any point before meeting a caustic, defines a {\it light sheet}. We now consider the entropy ${ S}_{\rm sheet}$ on this light sheet; this can also be defined by the entropy that `crosses' the light sheet from one side to the other. 

\bigskip

(e) There are two versions of the covariant entropy bond that we will study:

\b

(i) The covariant entropy bound originally proposed in \cite{bousso}  says that
\be
S_{\rm sheet}\le S_{\rm bound}
\label{bbound}
\ee
where
\be
S_{\rm bound}={A\over 4G}
\label{qbound}
\ee
with $A$ being the area of ${\cal S}$.

\b

(ii) A generalized version of this bound was proposed in \cite{fmw}.
In this version, we do not need to construct the light sheet all the way to the caustic; we can stop following the null rays at any point before they reach the caustic.  The end points of these null rays then describe a surface with area $A'$. Let $S_{\rm sheet}$ be the entropy that passes through the part of the light sheet that these null rays describe. Then the generalized  covariant entropy bound is the requirement (\ref{bbound}), with 
 \be
 S_{\rm bound}= {\Delta A\over 4G}
 \label{diff}
 \ee
 where
\be
\Delta A\equiv  A-A'
\label{diffp}
\ee
(Note that $A'<A$ because the light rays are converging along their path.)

 \bigskip
 
We will first look at the form (\ref{diff}) of the covariant entropy bound in the case where the surfaces with areas $A, A'$ are very close to each other; in this situation it will be easy to see that the equation of state (\ref{density}) can violate the bound for suitable initial conditions. Next we will consider complete evolutions where we follow the light sheet all the way to its caustic at the initial singularity; in this case the bound is given by (\ref{qbound}). The latter computation will allow us to address some efforts that were devoted to proving the covariant entropy bound under certain assumptions. 

\section{Violating the bound in an infinitesimal time slice}\label{infinitesimal}

In this section we will look at the violation of the covariant entropy bound in the simplest setting: where the light sheet is infinitesimally thin in the time direction. In this situation the metric in the neighbourhood of the sheet is given just by a choice of initial conditions.

\subsection{The metric}

We wish to consider a metric of the form encountered in a flat cosmology:
\be
ds^2=-dt^2+\sum_{i=1}^d a_i^2(t) dx_i dx_i
\label{metric}
\ee
We have in mind that the full theory of quantum gravity is given by M-theory, so we should have 10 space directions. But in what follows it will be equally easy to let the number of space dimensions be an arbitrary integer $d$. 

The light sheet we use will be confined to the interval
\be
t_0-\Delta t ~\le~ t~ \le~ t_0
\label{slice}
\ee
where we will take the limit of $\Delta t$ small. The metric is  subject to Einstein's equations, which are second order equations for the metric components. We can choose the $a_i, \dot a_i$ at time $t_0$:
\be
a_i(t_0)\equiv a_{i0}, ~~~~{\dot a_i(t_0)\over a_i(t_0)}\equiv b_{i0} 
\ee
subject only to the constraint set by the Einstein equation $G^t{}_t=8\pi G T^t{}_t$:
\be
-\h \left(\sum_i{{\dot a}_i\over a_i}\right)^2+\h \sum_i {{\dot a}_i^2\over a_i^2}=-8\pi G\rho
\label{gtt}
\ee
This constraint gives
\be
-\h \left(\sum_i b_{i0}\right)^2+\h \sum_i b_{i0}^2=-8\pi G \rho_0
\label{brho}
\ee
where $\rho_0=\rho(t_0)$. 
The $G^k{}_k$ equations for the space directions $k$ then determine the $\ddot a_k$ (there is no sum over $k$):
\bea
G^k{}_k&=&{\ddot a_k\over a_k} +{\dot a _k\over a_k}\(\sum_{i}{\dot a_i \over a_i}\)-{\dot a_k^2\over a_k^2}\nn
&&-\h \[2\sum_i {\ddot a_i\over a_i}+\(\sum_i{\dot a_i\over a_i}\)^2-\sum_i {\dot a_i^2\over a_i^2}\]\nn
&=&8\pi G T^k{}_k
\label{gkkl}
\eea
These equations have the form
\be
M \ddot a ={\cal S}
\label{ddota}
\ee
where $\ddot a$ is a column vector with entries $\{ \ddot a_1, \ddot a_2, \dots , \ddot a_d\}$, $M$ is the matrix with entries
\be
M_{ij}=\delta_{ij}-1
\ee
and ${\cal S}$ is a column vector with entries 
\be
{\cal S}_k=8\pi G T^k{}_k-{\dot a_k\over a_k}\(\sum_{i}{\dot a_i \over a_i}\)+{\dot a_k^2\over a_k^2}+\h \(\sum_i{\dot a_i\over a_i}\)^2+\sum_i {\dot a_i^2\over a_i^2}
\ee
We have
\be
\det M = (1-d) \ne 0
\ee 
so solving the relation (\ref{ddota}) yields finite values of the $\ddot a_k$. Thus in the interval (\ref{slice}) we get
\be
a_i(t)=a_{i0}[1+b_{i0} (t-t_0)] + O(\Delta t)^2
\label{ots}
\ee
Since we are interested in the limit $\Delta t \rightarrow 0$, we will just write
\be
a_i(t)=a_{i0}[1+b_{i0} (t-t_0)] , ~~~~~t_0-\Delta t~\le~ t~ \le~ t_0
\label{segment}
\ee
in what follows. We will take
\be
b_{i0}>0, ~~~~~~i=1, \dots, d
\label{bpos}
\ee
so that the slice $t_0-\Delta t <t<t_0$ represents a segment of an expanding cosmology. (We take $a_{i0}$ as positive numbers  as well.)

Note that we have used very little information about the matter content of the theory in constructing  the expanding segment (\ref{segment}). We can always choose the initial values of the metric and its first derivative, with the only constraint being (\ref{gtt}).\footnote{In the initial conditions we have taken the off-diagonal components of the metric to vanish; the corresponding Einstein equations tell us that these off diagonal components will be $O(\Delta t)^2$ in our slice, and so can be ignored in (\ref{segment}). In fact the Einstein equations have a $x^i \leftrightarrow -x^i$ symmetry. Thus assuming that  the matter stress tensor also has this symmetry, the off diagonal terms $g_{ti}$ and $g_{ij}, i\ne j$ will continue to vanish for all time if they are taken to vanish in the initial conditions.}

\subsection{Checking the covariant entropy bound}

Let us follow the steps (a)-(e) in section\,\ref{secbound} where we described the covariant entropy bound with the help of an illustration; we will use the metric and hypersurface etc. used in that illustration.  We label the steps below as (a')-(e'), in accordance with the corresponding steps in section \ref{secbound}.

\bigskip

(a') Our metric is given by (\ref{metric}),(\ref{segment}). Consider the $(d-1)$-dimensional  hypersurface ${\cal S}$ defined by (\ref{hyper}),(\ref{times}). The area of this hypersurface is
\be
A=\prod_{i=2}^d L_i a_i(t_0)=\prod_{i=2}^d L_i a_{i0}
\label{areaq}
\ee

\bigskip

(b') Consider a null geodesic starting at a point $\{x^2_0, \dots x^d_0\}$ on ${\cal S}$. The geodesic starts with a tangent vector which has with nonzero components
\be
{dx^1\over d\lambda }<0, ~~~~{dt\over d\lambda} <0
\label{direction}
\ee
so it heads to the past,   in the direction of decreasing   $x^1$, and in a direction normal to ${\cal S}$. The coordinate $x^1$ along the geodesic satisfies (\ref{dxdt}), which gives, in the infinitesimal slice (\ref{slice})
\be
x^1(t)=x^1_0-{(t_0-t)\over a_{10}}
\label{xone}
\ee

\bigskip

(c') We require that the set of null geodesics constructed this way be converging as we move away from ${\cal S}$. The condition for this is (\ref{convergence}). We find
\be
{dA\over dt}= {d\over dt} \left (\prod_{i=2}^d L_ia_i(t)\right ) =\left (\prod_{i=2}^d L_ia_{i0} \right ) \left ( \sum_{j=2}^d b_{i0}\right ) 
  \ee
  where we have used (\ref{segment}) and kept only terms that are nonzero in the limit $\Delta t\rightarrow 0$. From (\ref{bpos}) we find
 \be
{dA\over dt}>0
\ee
 in accordance with the requirement (\ref{convergence}). 

\bigskip

(d') Our light sheet is made of the above null rays, followed back from the surface $t=t_0$ to the surface $t=t_0-\Delta t$. 

\begin{figure}[h]
\includegraphics[scale=.82]{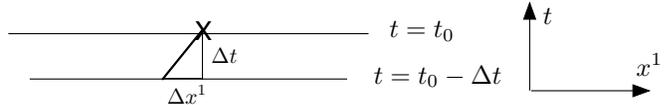}
\caption{\label{fsheet2pre} The same set-up as in fig.\ref{fsheet}, for a light sheet that lies between two spacelike hypersurfaces separated by an infinitesimal distance $\Delta t$.   The entropy on the spatial segment $\Delta x^1$ will pass through the light sheet in the time $\Delta t$.}
\end{figure}

Let us now compute the entropy passing through this light sheet.  On the spatial slice at this time $t$, consider the sliver of space given by
\be
x^1_0-\Delta x^1 \, \le  x^1 \,\le  x^1_0, ~~~~0\, \le x^i\, \le L_i, ~~~~i=2, \dots, d
\ee
where
\be
\Delta x^1={\Delta t\over a_{10}} 
\ee
As can be seen from fig.\ref{fsheet2pre}, the entropy passing through our light sheet is equal to the entropy present on this sliver of space.  The proper volume of this sliver is
\be
\Delta V=a_{10} \Delta x^1 \left (\prod_{i=2}^d a_{i0} L_i\right )= \Delta t \left (\prod_{i=2}^d a_{i0} L_i\right )
\ee
The entropy density on this sliver is
\be
s=K\sqrt{\rho_0\over G}
\ee
 Thus the entropy $S_{\rm sheet}$ passing through our light sheet, which equals the entropy on the sliver,  is
\be
S_{\rm sheet} = s \Delta V = K\sqrt{\rho_0\over G}\, \Delta t \prod_{i=2}^d a_{i0} L_i
\label{ssheet}
\ee

\bigskip

(e') We will use the form (\ref{diff}) of the covariant entropy bound.   The area $A$ is given by (\ref{areaq}). The area $A'$ is the area of the surface at the lower end of the light sheet. We have
\bea
A'&=&\prod_{i=2}^d L_i a_i(t_0-\Delta t)=\prod_{i=2}^d L_i a_{i0}(1-b_{i0}\Delta t)\nn
&=&\left( \prod_{i=2}^d L_i a_{i0}\right ) \left ( 1-\Delta t\sum_{i=2}^d b_{i0}\right )
\label{areaeqp}
\eea
Thus
\be
S_{\rm bound}={A-A'\over 4G}={\left( \prod_{i=2}^d L_i a_{i0}\right )\Delta t \sum_{i=2}^d b_{i0}\over 4G}
\ee
We find that
\be
r\equiv {S_{\rm sheet}\over S_{\rm bound}}={4K\sqrt{\rho_0 G} \over \sum_{i=2}^d b_{i0} }
\label{req}
\ee
Substituting $\rho_0$ from (\ref{brho}), we get
\be
r={K\over \sqrt{\pi}}{ \[  \(\sum_{i=1}^d b_{i0}\)^2 -\sum_{i=1}^d b_{i0}^2\]^\h\over \sum_{i=2}^d b_{i0} }
\ee

It is easy to see that for any given value of $K$, we can choose the $b_{i0}$ to get $r>1$, thus violating the covariant entropy bound. For example suppose we set 
\be
b_{i0}=b, ~~~~ i=2, \dots d
\label{fix1}
\ee
We then find that   $r>1$ if we take
\be
{b_{10}\over b} ~>~ \frac{\pi(d-1)}{2K^2} - \frac{d-2}{2}
\label{bb}
\ee
The relation (\ref{bb}) is the main result of this paper. This inequality shows that we can violate the covariant entropy bound by an asymmetric choice of initial conditions on a spacelike slice, if the entropy density is given by (\ref{density}). On the other hand it has ben argued that an entropy density of the form (\ref{density}) can be obtained by a collection of string states at the Horowitz-Polchinski correspondence point \cite{veneziano}, or by a gas of black holes \cite{bf}, or by a collection of intersecting brane states that describe black holes in string theory \cite{alimathur}. Putting this fact together with the inequality (\ref{bb}), we find a violation of the covariant entropy bound.

In section\,\ref{discussiona} we will discuss if it is possible to avoid this conclusion; for example there may be a physical reason why asymmetric expansion like (\ref{bb}) is not allowed. We do not however know of any such clear reason, and looking for one may open up interesting avenues of exploration.

\section{The full cosmology for the equation of state $p=\rho$}\label{full}

We have seen that using a thin slice (\ref{segment}) of a flat cosmology was enough to yield a violation of the covariant entropy bound for the equation of state (\ref{density}). Nevertheless, it is helpful to get a perspective of the full evolution under this equation of state, and to see the violation of the bound in this situation. In this section we will consider the evolution assuming a perfect fluid with $p=\rho$; in the next section we will consider the effect of viscosity on the evolution. 

\subsection{The metric for the equation of state $p=\rho$ }\label{secfour}

In this section we will  solve the Einstein equations for our flat cosmology with equation of state $p=\rho$ (eq. (\ref{eos})), assuming the stress tensor of  a perfect fluid.

 The metric is assumed to have the form
\be
ds^2=-dt^2+\sum_{i=1}^d a_i^2(t) dx_i dx_i
\label{metric2}
\ee
Suppose the pressure in the direction $x_i$ is given by $p_i=w_i \rho$. 
The $G^{0}{}_0$ component of Einstein's equations gives the Friedmann equation
\be
-\h \left(\sum_i{{\dot a}_i\over a_i}\right)^2+\h \sum_i {{\dot a}_i^2\over a_i^2}=-8\pi G\rho
\label{gtt2}
\ee
The $G^k{}_{k}$ components give
\bea
{\ddot a_k\over a_k} +{{\dot a}_k\over a_k}\left(\sum_{i}{{\dot a}_i \over a_i}\right)&-&{{\dot a}_k^2\over a_k^2}- \sum_i {\ddot a_i\over a_i}\nonumber \\
&=&8\pi G (1+w_k)\rho=16\pi G \rho
\label{gkkpre2}
\eea
where in the second step we have set $w_i=1$ for all $i$ in accordance with our equation of state (\ref{eos}).\footnote{The solution for general $w_i$ was found in \cite{cm}.} The above equations can be solved with the ansatz
\be
a_i=a_{0i} t^{C_i}, ~~~i=1, \dots d
\label{ai}
\ee
with $a_{0i}, C_i$ constants. Then
\be
{\dot a_i\over a_i}={C_i\over t}, ~~~~~~
{\ddot a_i\over a_i}={C_i (C_i-1)\over t^2}
\ee
Equations (\ref{gtt2}) and (\ref{gkkpre2}) give respectively
\be
{1\over t^2} \left [ -\h \(\sum_i C_i\)^2+\h \sum_i C_i^2\right ] =-8\pi G \rho
\label{eqone2}
\ee
\bea
&&{1\over t^2} \left [ C_k (C_k-1)+C_k\sum_i C_i- C_k^2-\sum_i C_i(C_i-1)\right ] \nn
&&\qquad\qquad\qquad\qquad\qquad\qquad= 16\pi G\rho
\label{eqtwo2}
\eea
Eliminating $\rho$ gives 
\be
C_k\(\sum_i C_i-1\)=\sum_j C_j \( \sum_i C_i-1\)
\label{ckk2}
\ee
There are two cases:

\bigskip

(1) Suppose
\be
\sum_i C_i \ne 1
\ee
Then we get
\be
C_k=\sum_i C_i\equiv C_T
\label{ck2}
\ee
Summing over the $d$ possibilities for $k$ gives $d \, C_T= d^2 C_T$, which implies $C_T=0$. From (\ref{ck2}), we get
\be
C_k=0, ~~~k=1, \dots, d
\ee
From (\ref{gtt2}) we find $\rho=0$, so we just have empty, toroidally compactified,  Minkowski space. This case will not be of  interest to us.

\bigskip

(2) The other option is
\be
\sum_i C_i=1
\label{condition}
\ee
Then the relations (\ref{ckk2}) place no constraints on the $C_k$ apart from the condition (\ref{condition}). From (\ref{eqone2}) we get
\be
\rho={1\over 16\pi G t^2}\left(1-\sum_i C_i^2\right)
\label{qqtwo}
\ee
To get $\rho>0$ we need
\be
\sum_i C_i^2 <1
\label{qqone}
\ee
We will take this solution for the metric and examine the entropy bound using the entropy density (\ref{density}).

\subsection{Checking the bound}\label{secchecka}

Let us follow the steps (a)-(e) in section\,\ref{secbound} where we described the covariant entropy bound with the help of an illustration; we will use the metric and hypersurface etc. used in that illustration.

Our metric is (\ref{metricq}). The scale factors evolve as
\be
a_i=a_{i0} t^{C_i}, ~~~i=1, \dots,  d
\label{avalues}
\ee
where $a_{i0}>0$ are constants and the $C_i$ satisfy (\ref{condition}),(\ref{qqone}). We assume in addition that
\be
C_i>0, ~~~i=1, \dots d
\label{cpositive}
\ee
so that all spatial directions collapse to zero size at the initial singularity $t=0$.

We again label the steps below as (a')-(e'), in accordance with the corresponding steps in section \ref{secbound}.

\bigskip

(a') Consider the $(d-1)$-dimensional  hypersurface ${\cal S}$ defined by (\ref{hyper}),(\ref{times}). The area of this hypersurface is
\be
A=\prod_{i=2}^d L_i a_i(t_0)=\prod_{i=2}^d L_i a_{i0}t_0^{C_i}
\label{areaeq2}
\ee

\bigskip

(b') Consider a null geodesic starting at a point $\{x^2_0, \dots x^d_0\}$ on ${\cal S}$. The geodesic starts with a tangent vector
\be
{dx^1\over d\lambda }<0, ~~~~{dt\over d\lambda} <0
\label{direction2}
\ee
so it heads to the past,   in the direction of decreasing   $x^1$, and in a direction normal to ${\cal S}$. The coordinate $x^1$ along the geodesic satisfies (\ref{dxdt}), which gives
\be
x^1(t)=x^1_0-\int_{t=t}^{t_0} {dt\over a_{10}t^{C_1}}=x^1_0-{t_0^{1-C_1}-t^{1-C_1}\over a_{10}(1-C_1)}
\label{xone2}
\ee

\bigskip

(c') We require that the set of null geodesics constructed this way be converging as we move away from ${\cal S}$. The condition for this is (\ref{convergence}). We find
\be
{dA\over dt}=\left ( \prod_{i=2}^d a_{i0} \right ){d\over dt}\left ( t^{\sum_{I=2}^d C_i}\right )= \left ( \prod_{i=2}^d a_{i0}\right ){d\over dt} t^{1-C_1}
  \ee
 where in the last step we have used (\ref{condition}). From the condition (\ref{qqone}) we find that 
$ 1-C_1>0$. This gives ${dA/ dt}>0$, in accordance with the requirement (\ref{convergence}). 

\bigskip

(d') It is easy to see that as we follow the null geodesics backwards in time, the separation between neighboring  geodesics reaches zero at $t=0$, but not before. The  surface formed by these geodesics from their start at ${\cal S}$ to their endpoint at the singularity $t=0$ is thus a `light sheet'.

\begin{figure}[h]
\includegraphics[scale=.82]{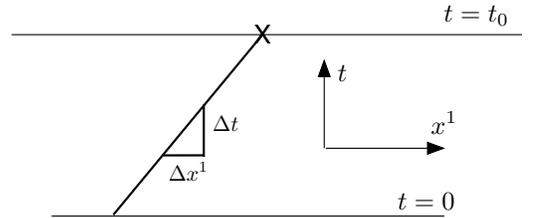}
\caption{\label{fsheet2} The same set-up as in fig.\ref{fsheet}. We consider the entropy passing through a  small part of the light sheet. The entropy on the spatial segment $\Delta x^1$ will pass through the light sheet in the time $\Delta t$.}
\end{figure}

Let us now compute the entropy passing through this light sheet. Take a time $t$, with $0<t<t_0$. The position of the light sheet at this time is given by the value of $x^1(t)$ given in (\ref{xone2}). On the spatial slice at this time $t$, consider the sliver of space given by
\be
x^1(t)-\Delta x^1 \, \le  x^1 \,\le  x^1(t), ~~~~0\, \le x^i\, \le L_i, ~~~~i=2, \dots, d
\ee
As can be seen from fig.\ref{fsheet2}, the entropy on this sliver of space will soon thereafter pass through the light sheet. The spatial volume of this sliver is
\be
\Delta V=a_1(t) \Delta x^1 \prod_{i=2}^d a_i(t) L_i
\ee
The entropy density on this sliver is
\be
s=K\sqrt{\rho\over G}=KG^{-\h}\left[{\left(1-\sum_i C_i^2\right)\over 16\pi G t^2}\right]^\h
\ee
where we have used (\ref{qqtwo}). Thus the entropy on the sliver is
\bea
\Delta s &=& s \Delta V\nn
&=& KG^{-\h}\left[{\left(1-\sum_i C_i^2\right)\over 16\pi G t^2}\right]^\h a_1(t) \Delta x^1 \prod_{i=2}^d a_i(t) L_i\nn
\eea
We can write $\Delta x^1 = \Delta t/a_1(t)$, where $\Delta t$ is the time over which the entropy on the sliver crosses the light sheet (fig.\ref{fsheet2}). This gives
\bea
\Delta s &=& KG^{-\h}\left[{\left(1-\sum_i C_i^2\right)\over 16\pi G t^2}\right]^\h \Delta t  \prod_{i=2}^d a_i(t) L_i\nn
\eea
Adding over all such intervals $\Delta t$ gives the total entropy $S_{\rm sheet}$ passing through our light sheet. Using (\ref{avalues}) we find
\bea
S_{\rm sheet}&=&{K\prod_{i=2}^d a_{i0}L_i\over G}\left[{1-\sum_i C_i^2\over 16\pi }\right]^\h\int_{t=0}^{t_0} dt {t^{\sum_{i=2}^d C_i}\over t}\nn
&=&{K\prod_{i=2}^d a_{i0}L_i\over G}\left[{1-\sum_i C_i^2\over 16\pi }\right]^\h {t_0^{\sum_{i=2}^d C_i}\over \sum_{i=2}^d C_i}\nn
&=&{K\prod_{i=2}^d a_{i0}L_i\over G}\left[{1-\sum_i C_i^2\over 16\pi }\right]^\h {t_0^{1-C_1}\over 1-C_1}
\label{ssheet2}
\eea
where in the last step we have used (\ref{condition}).

\bigskip

(e') We consider the covariant entropy bound  in the form  given by eq.(\ref{qbound}). Using (\ref{areaeq2}) we find
\bea
S_{\rm bound}&=&{A\over 4G}={1\over 4G}\left ({\prod_{i=2}^d L_i a_{i0}}\right )t_0^{\sum_{i=2}^d C_i}\nn
&=&{1\over 4G}\left ({\prod_{i=2}^d L_i a_{i0}}\right )t_0^{1-C_1}
\eea
We find that
\be
r\equiv {S_{\rm sheet}\over S_{\rm bound}}={K\left(1-\sum_i C_i^2\right)^\h\over \sqrt{\pi}(1-C_1)}
\label{req2}
\ee
It is easy to see that for any given value of $K$, we can choose the $C_i$ to get $r>1$, thus violating the covariant entropy bound. For example suppose we set 
\be
C_i=\t C, \, i=2, \dots d
\label{fix12}
\ee
 Using (\ref{condition}) we can write $\t C$ in terms of $C_1$
 \be
 C_1=1-(d-1)\t C
 \label{fix2}
 \ee
  We then find that   $r>1$ if we take
\be
1-{{2K^2\over \pi}\over 1+{K^2 d\over \pi (d-1)}}<C_1<1
\label{c1bound}
\ee
where the last inequality follows from (\ref{condition}) and (\ref{cpositive}). It can be checked that if we take a thin slice of the full geometry studied in this section, then the condition (\ref{c1bound}) is consistent with the general condition (\ref{bb}).

\subsection{Comments on the computation}

Let us now make a few comments on the above computation.

\bigskip

(a) In the above computation we had used an open surface ${\cal S}$. We can perform a similar check for a closed surface ${\cal S}$, as follows. The open surface ${\cal S}$ was defined by the coordinate intervals
\be
0\le x^i\le L_i, ~~~i=2, \dots, d
\ee
Let us compactify spacetime so that this cuboid becomes a torus $T^{d-1}$
\be
(x^i=0) ~~\sim ~~ (x^i=L_i), ~~~~~~~i=2, \dots, d
\ee
In (\ref{xone2}) choose $x^1_0$ such that 
\be
x^1_0={t_0^{1-C_1}\over a_{10}(1-C_1)}
\ee
With this choice, the null geodesics  starting on ${\cal S}$ reach $x^1=0$ at $t=0$. 

\begin{figure}[h]
\includegraphics[scale=.82]{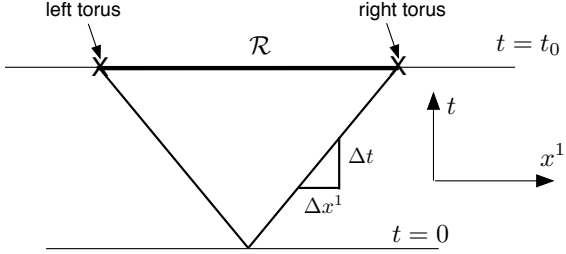}
\caption{\label{fsheet3} The region ${\cal R}$; its boundaries are marked by the crosses, which depict compact torii. We compute the entropy through the light sheets indicated, and compare this to the total area of the two torii.}
\end{figure}

\begin{figure}[h]
\includegraphics[scale=.38]{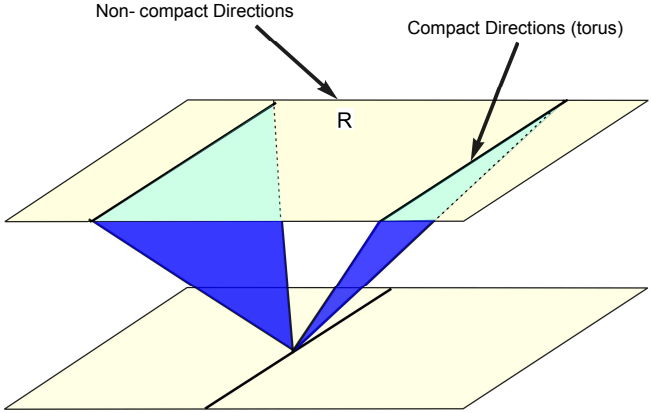}
\caption{\label{fcosmo2} The same as in fig.\ref{fsheet3}, shown in a 3-dimensional depiction. The light rays making up the light sheet maintain a constant coordinate position in the compact directions, while moving radially in the noncompact directions.}
\end{figure}

Now consider the $(d-1)$-dimensional region ${\cal R}$ described as follows
\be
t=t_0, ~~~ -x^1_0\le x^1\le x^1_0, ~~~0\le x^i\le L_i, ~~i=2, \dots, d
\ee
The $x^1$ direction gives a 1-dimensional line segment with two endpoints; at each endpoint we have the torus $T^{d-2}$ described by the compact directions $x^i,\, i=2, \dots. d$. These two $T^{d-2}$-dimensional torii form the boundary of the region ${\cal R}$, and their union is the surface ${\cal S}$ that will appear in the Bousso bound. The light sheets starting at each of these torii are constructed just as in section\,\ref{secchecka}: the null geodesics stay at a constant value of $x^i, ~~i=2, \dots, d$, while moving in the $x^1$ direction as
\be
{dx^1\over dt}=\pm {1\over a_1(t)}
\ee
(The two signs are for the right and left torii respectively.) We thus get the wedge shape depicted in figs.\ref{fsheet3},\ref{fcosmo2}. The entropy through each of the slanted portions of the wedge is equal to the value $S_{\rm sheet}$ found in (\ref{ssheet2}). The area of each torus is equal to the area ${\cal A}$ given in (\ref{areaeq2}). Thus for the present case both $S_{\rm sheet}$ and $S_{\rm bound}$ are twice the values in (\ref{req2}), and we get the same value for their ratio $r$ as before
\be
r\equiv {S_{\rm sheet}\over S_{\rm bound}}={K(1-\sum_i C_i^2)^\h\over \sqrt{\pi}(1-C_1)}
\ee
Thus the choice (\ref{c1bound}) will again violate the bound.


In the above example, the surface ${\cal S}$ was compact, since it was the boundary of a region ${\cal R}$. The region ${\cal R}$ had  one noncompact direction, while its other directions were compact. It does not seem possible to violate the bound (for an arbitrary value of $K$)  if we choose a region ${\cal R}$ which has more than one noncompact direction. It would be interesting to understand the implications of this restriction better.

\bigskip

(b) The equation of state (\ref{density}) has $s\sim \rho^\h$. If one considers the cases $s\sim \rho^\mu$ with $\mu>\h$, then we do not get a violation of the bound. In this latter case one finds that the ratio $r={S_{\rm sheet}/S_{\rm bound}}$ decreases with $t$; we reproduce the relevant computation in the appendix \ref{Appendix1}. Thus if the bound is satisfied at the Planck time, then it will be satisfied at later times. For the case $\mu=\h$, the ratio $r$ becomes time independent, and then the choice (\ref{c1bound}) violates the bound at all $t$ where this equation of state holds. 

 \section{The effect of viscosity}\label{viscous}
  
  We have seen that the equation of state (\ref{density}) gives $p=\rho$. But it is possible that the fluid has in addition a viscosity, which will add further terms to the  stress tensor. In this section we will consider how the evolution of the last section is modified by viscous effects.\footnote{We thank Soo-Jong Rey for pointing out the possible connections between entropy bounds and the viscosity bound.}
  
  With viscosity $\eta$, the stress tensor $T_{\mu\nu}$ gets an additional term $-2\eta\sigma_{\mu\nu}$, where $\sigma_{\mu\nu}$ is the traceless symmetric part of the velocity gradients. With a metric of the form (\ref{metric}), we get
  \bea
  T^0{}_0&=&-\rho\nn
T^k{}_k&=&p-2\eta \sigma^k{}_k=p-2\eta\[{\dot a_k\over a_k}-{1\over d}\sum_{l=1}^d {\dot a_l\over a_l}\]\nn
&=&\rho-2\eta\[{\dot a_k\over a_k}-{1\over d}\sum_{l=1}^d {\dot a_l\over a_l}\]
\eea
   where in the last step we have set $p=\rho$. 
 
   We do not know the value of $\eta$ for the `black hole gas', but it was conjectured in \cite{son} that as a general rule
   \be
   {\eta\over s}\ge {1\over 4\pi}
   \label{son}
   \ee
   This inequality was derived using AdS/CFT, where a black hole yielded ${\eta/ s}={1/4\pi}$; conjecturing that a black hole has the lowest ${\eta/ s}$ yields (\ref{son}). Let us conjecture that the black hole gas also has this minimum viscosity, so that
   \be
   \eta={1\over 4\pi} s={K\over 4\pi} \sqrt{\rho\over G}
   \ee
   Now let us consider the Einstein equations. The $G^{0}{}_0$ component gives the Friedmann equation
\be
-\h \left(\sum_i{{\dot a}_i\over a_i}\right)^2+\h \sum_i {{\dot a}_i^2\over a_i^2}=8\pi G T^0{}_0=-8\pi G\rho
\label{gtt3}
\ee
The $G^k{}_{k}$ components give
\bea
{\ddot a_k\over a_k} &+&{{\dot a}_k\over a_k}\left(\sum_{i}{{\dot a}_i \over a_i}\right)-{{\dot a}_k^2\over a_k^2}- \sum_i {\ddot a_i\over a_i}\nonumber \\
&=&-8 \pi G T^0{}_0+8\pi G T^k{}_k\nn
&=&8\pi G \[\rho + \rho-{K\over 2\pi}\sqrt{\rho\over G}\({\dot a_k\over a_k}-{1\over d}\sum_{l=1}^d {\dot a_l\over a_l}\)\]\nn
&=&8\pi G \[2 \rho-{K\over 2\pi}\sqrt{\rho\over G}\({\dot a_k\over a_k}-{1\over d}\sum_{l=1}^d {\dot a_l\over a_l}\)\]\nn
\label{gkkpre3}
\eea
With dissipation, we no longer expect to get the simple power law solutions that we had without dissipation. But it is instructive to check this fact explicitly. Thus we assume the ansatz
\be
a_i=a_{0i} t^{C_i}, ~~~i=1, \dots d
\label{ai3}
\ee
with $a_{0i}, C_i$ constants. Equations (\ref{gtt3}) gives
\be
{1\over t^2} \left [ -\h \(\sum_i C_i\)^2+\h \sum_i C_i^2\right ] =-8\pi G \rho
\label{eqone3}
\ee
Eq. (\ref{gkkpre3}) gives
\bea
&&{1\over t^2} \left [ C_k (C_k-1)+C_k\sum_i C_i- C_k^2-\sum_i C_i(C_i-1)\right ] \nn
&&\qquad\qquad\qquad= 8\pi G \[2 \rho-{K\over 2\pi}\sqrt{\rho\over G}{1\over t}\(C_k-{1\over d}\sum_{l=1}^dC_l\)\]\nn
\label{eqtwo3}
\eea
We can find $\rho$ from (\ref{eqone3}) and substitute into (\ref{eqtwo3}); this gives
\bea
&&\!\!\!\!\!\!\!\!\!\!\!\!{1\over t^2} \left [ C_k (C_k-1)+C_k\sum_i C_i- C_k^2-\sum_i C_i(C_i-1)\right ]\nn
&=&
 8\pi G \[2 \rho-{K\over 2\pi}\sqrt{\rho\over G}{1\over t}\(C_k-{1\over d}\sum_{l=1}^dC_l\)\]\nn
  &=& {1\over t^2}  \[ \(\sum_i C_i\)^2-\sum_i C_i^2 \]\nn &-&  {K\sqrt{2\over \pi}} {1\over t^2}
  \[  \h \(\sum_i C_i\)^2-\h \sum_i C_i^2\]^\h \(C_k-{1\over d}\sum_{l=1}^dC_l\)\nn
  \label{equalterms3}
\eea
 Solving for the $C_k$ gives
 \be
C_k=\(\sum_i C_i\)\times {\(\sum_i C_i-1\) + {K\over d\sqrt{\pi}} \[ \(\sum_i C_i\)^2-\sum_i C_i^2  \]^\h  \over \(\sum_i C_i-1\)+{K\over \sqrt{\pi}} \[   \(\sum_i C_i\)^2- \sum_i C_i^2  \]^\h}
\label{finaleq3}
\ee
so that all the $C_k$ are equal. Summing (\ref{finaleq3}) over $k$ gives (after a little algebra), the relation
 \be
 \sum_i C_i=1
 \ee
so that we get
\be
C_k={1\over d}, ~~~~k=1, \dots , d
\label{ckk3}
\ee
Thus the only solution in the power law ansatz is an isotropic one. For an isotropic expansion the viscosity term has no effect, and (\ref{ckk3}) agrees with the isotropic solution allowed by (\ref{condition}) in the zero viscosity case. 

Even though we have not found the general solution in the case of nonzero viscosity, the lesson we wish to draw can already be seen from (\ref{equalterms3}). We expect that if we start from anisotropic initial conditions, then the solution will eventually asymptote to the isotropic power law solution (\ref{ckk3}). From (\ref{equalterms3}) we see that the contribution of the terms from the geometry is of the same order as the terms from the viscous effects. Thus a solution that is initially anisotropic will isotropize on a timescale that is of the order of the Hubble scale $H^{-1}$; in particular this isotropization will not happen `instantaneously' or over planck times. Thus even in the presence of viscosity, we can return to the computation of section \ref{infinitesimal} where we set up anisotropic initial conditions, and follow the evolution for a small time $\Delta t$; viscous effects will contribute only to the $O(\Delta t)^2$ terms in (\ref{ots}), and will thus not affect the argument showing violation of the generalized covariant entropy bound.

\section{The entropy bound based on the apparent horizon}\label{apparent}
 
 Let us also consider the entropy bound proposed in \cite{br}. Due to the expansion of the cosmology, we can find spacelike surfaces of dimension $(d-1)$  that form an apparent horizon. Take any such surface, and let the area of this apparent horizon be $A$. Then the conjecture states that the entropy within this apparent horizon will be bounded by $A/4G$.

 To get a violation of the bound for the equation of state (\ref{density}), we will again need to have an asymmetric expansion; i.e., not all the $C_i$ in (\ref{ai}) would be the same.  In this situation the apparent horizon is not  a spherical surface, making  computations difficult. To avoid this problem, we take a spacetime where  some of the directions are compactified to circles, while the others are noncompact. We will then let these two sets of directions expand at different rates, getting the required asymmetry.
 
We let $d_1$ directions be noncompact and $d_2=d-d_1$ directions be compact. Let each compact direction have a coordinate range
\be
0\le x^i\le L, ~~~~i=d_1+1, \dots, d
\ee
We let
\bea
a_i&=& a_0 t^C\equiv a_{nc}(t), ~~~~i=1, \dots, d_1\nn
a_i&=& \t a_0 t^{\t C}\equiv a_c(t), ~~~i=d_1+1, \dots d
\label{evolve}
\eea
with 
\be
C>0, ~~\t C>0
\ee
The condition (\ref{condition}) gives
\be
d_1 C + d_2 \t C =1
\label{conditionp}
\ee

 The metric (\ref{metric})  in the noncompact directions is now homogenous and isotropic. We can thus write it in the form
 \be
 ds^2=-dt^2+a^2_{nc}(t) \(dr^2+r^2 d\Omega^2\)
\ee
 Following \cite{br}, we define $\t r = a_{nc} r$, and compute $\nabla \t r$ in the directions $t,r$
 \be
 \nabla \t r = \{ \t r _{,t}, \t r _{, r}\} = \{ \dot a_{nc} r, a_{nc} \}
 \ee
 We then have
 \be
\left|\nabla \t r\right|^2 = 
\t r_{,\mu} \t r_{,\nu} g^{\mu\nu}= 1-\dot a^2_{nc} r^2
\ee
This vanishes when
\be
r={1\over \dot a_{nc}}
\ee
This corresponds to a proper radius of the apparent horizon
\be
X_a= a_{nc} r = {a_{nc}\over \dot a_{nc}}, 
\ee
a value that equals the Hubble radius of the expansion. Noting from (\ref{evolve}) that $a_{nc}=a_0 t^C$, we get
\be
X_a={t\over C}
\ee
The volume $V_c$ of the compact directions is 
\be
V_c=V_0 t^{d_2 \t C}
\label{vcompact}
\ee
where $V_0=(\t a_0 L)^{d_2}$.  
 The area of the apparent horizon (including the extent in the compact directions) is
\be
A=\Omega_{d_1-1} X_a^{d_1-1}V_c=\Omega_{d_1-1} \({t\over C}\)^{d_1-1}V_0 t^{d_2 \t C}
\ee
The volume of the noncompact directions is
\be
V_{nc}={\Omega_{d_1-1}\over d_1} X^{d_1}={\Omega_{d_1-1}\over d_1} \({t\over C}\)^{d_1}
\ee
 Thus the overall volume inside the apparent horizon  is
\be
V={\Omega_{d_1-1}\over d_1} \({t\over C}\)^{d_1}~V_0 t^{d_2 \t C}
\ee
 From (\ref{qqtwo}) we have
\be
\rho={1\over 16\pi G t^2} \(1-d_1 C^2 -d_2 \t C^2\)
\label{qqtwoa}
\ee
This gives, with the entropy expression (\ref{oneqq})
\bea
S = K \sqrt{{1\over 16\pi G^2 t^2} \(1-d_1 C^2 -d_2  \t C^2\)}  \nonumber \\
 \times {\Omega_{d_1-1}\over d_1} \({t\over C}\)^{d_1}V_0 t^{d_2 \t C}
\eea
We can write this as
\be
S=K\sqrt{(1-d_1 C^2 -d_2  \t C^2)\over \pi d_1^2 C^2}  ~ {A\over 4G}
\ee
Thus
\be
r={S\over S_{\rm bound}}=K\sqrt{(1-d_1 C^2 -d_2  \t C^2)\over \pi d_1^2 C^2}
\ee
Let us choose any $d_1\ge 1$ and any $d_2\ge 1$. The relation  (\ref{conditionp}) gives
\be
\t C= {1-d_1 C\over d_2}
\ee
 Then we can see that we get $r>1$ if we take
 \be
 C ~<~ {d_1 K^2+\sqrt{d_1 d_2 K^2 [(d-1)K^2+d_1(d_2-1)\pi]}\over d_1 (d \, K^2+d_1d_2\pi)}
 \ee
 where $d=d_1+d_2$.

  Thus we see that the entropy expression (\ref{density}) can violate the bound given in terms on the area of the apparent horizon.

\section{Relation to proofs of the Bousso's bound} \label{marolf}

There have been several approaches to proving the Bousso bound. These proofs start with some assumptions, and prove the bound subject to these assumptons. We will now look at some of these proofs, and see that the assumptions taken there do not hold for the equation of state and initial conditions on expansion that we have used to show a violation of the bound.

\subsection{The first proof of \cite{fmw} }

 In \cite{fmw} it was shown that under some assumptions on the matter stress tensor, one can actually prove the covariant entropy bound. Two such proofs were given. For the first proof, it was assumed  that the entropy current $s^a$ is constrained by the stress tensor $T_{ab}$ in the manner
 \be
 |s_a k^a|\le \pi(\lambda_\infty-\lambda)T_{ab} k^ak^b
 \label{wcond4}
 \ee
 Here $\lambda$ is an affine parameter along the null geodesics used in the covariant entropy  bound. We have $\lambda=0$ at the surface ${\cal S}$ used in the bound, and $\lambda=\lambda_\infty$ at the endpoint of the geodesic.
 
 For our example, we have the metric (\ref{metric}) and a null  geodesic moving in the plane $t, x^1$.  The null nature of the geodesic gives
 \be
 {dt\over d\lambda}=a_1 {dx^1\over d\lambda}
 \label{tx1}
 \ee
 The geodesic equation gives
 \be
 {d^2x^1\over d\lambda^2}=-2\Gamma^1_{10}{dx^1\over d\lambda}{dt\over d\lambda}=-2{\dot a_1 \over a_1}{dx^1\over d\lambda}{dt\over d\lambda}=-2{da_1\over d\lambda}{1\over a_1} {dx^1\over d\lambda}
 \ee
This gives
\be
{dx^1\over d\lambda}={\alpha\over a_1^2}={\alpha\over a_{10}^2} t^{-2 C_1}
\ee
where $\alpha<0$ due to  (\ref{direction}). We then have from (\ref{tx1})
\be
{dt\over d\lambda}={a_1}{\alpha\over a_{10}^2} t^{-2 C_1}={\alpha\over a_{10}} t^{- C_1}
\ee
which gives
\be
\lambda-\lambda_\infty=\int_{t=0}^t dt {a_{10}\over \alpha}t^{C_1}={a_{10}\over \alpha (1+C_1)} t^{1+C_1}
\ee
We have
\be \label{eq:EntropyDensity}
s^a=\{s, 0, \dots 0\}
\ee
and
\be
k^a=\left\{ {dt\over d\lambda}, {dx^1\over d\lambda}, 0, \dots, 0\right\}
\ee
Thus we have
\bea
|s_a k^a|&=&\left|s{dt\over d\lambda}\right|=K\sqrt{\rho\over G}{(-\alpha)\over a_{10}}t^{-C_1}\nn
&=&{K\over4G\pi^\h}(1-\sum_i C_i^2)^\h{(-\alpha)\over a_{10}}t^{-C_1-1}
\eea
With the equation of state $p=\rho$ (eq.(\ref{eos})) we have
\be \label{eq:StressTensor}
T_{ab}=\rho~ {\rm diag}\left\{ 1, a_1^2, \dots, a_d^2\right\}
\ee
Thus we get
\bea
T_{ab}k^ak^b&=&\rho \left ( {dt\over d\lambda}\right )^2 + \rho a_1^2 \left ( {d x^1\over d\lambda}\right )^2\nonumber \\&=&{(1-\sum_i C_i^2)\over 8\pi G t^2} {\alpha^2\over a_{10}^2}t^{-2C_1}
\eea
Then the inequality (\ref{wcond4}) becomes
\be
{K\over4G\pi^\h}\(1-\sum_i C_i^2\)^\h\le {1\over (1+C_1)} {\(1-\sum_i C_i^2\)\over 8 G } 
\ee
which simplifies to
\be
{K}\le {1\over (1+C_1)} {\pi^\h(1-\sum_i C_i^2)^\h\over 2  } 
\ee
We can write this as
\be
r'\equiv {2 K(1+C_1)\over \pi^\h (1-\sum_i C_i^2)^\h}\le 1
\ee
 We have
\be
{r'\over r}={2(1-C_1^2)\over \(1-\sum_i C_i^2\)}
\ee
where $r$ is given in (\ref{req}). Since $(1-\sum_i C_i^2)\le  (1- C_1^2)$, we find
\be
{r'\over r}\ge 2
\ee
Thus if we take initial conditions that give $r>1$ to violate the covariant entropy bound, then we would also have $r'>1$, so  we would violate the condition (\ref{wcond4}) of the proof of \cite{fmw}. Thus there is no conflict between the violation we find and the first proof of \cite{fmw}.

\subsection{The second proof of \cite{fmw}}

 In the second proof of \cite{fmw}, there are two assumptions about the matter content. We will now examine these, and find that the second condition is in fact violated by our equation of state and the chosen initial conditions. 
 
 The two assumptions are
\be
{\rm (i)} \qquad (s_a k^a)^2\le {\alpha_1\over   G}T_{ab}k^ak^b
\label{wcond1}
\ee 
\be
{\rm (ii)} \qquad |k^ak^b s_{a;b}|\le \alpha_2T_{ab}k^ak^b
\label{wcond2}
\ee
where $k^a$ is any null vector. The constants $\alpha_1, \alpha_2$ are constrained by 
\be
\left({\pi\alpha_1}\right)^{1\over 4}+\left({\alpha_2\over \pi}\right)^\h=1
\label{relationa}
\ee
Let us examine the conditions (i) and (ii) in turn.

\bigskip

(i) Our metric is  (\ref{metric}). 
 The entropy current  and the stress tensor are given respectively in \eqref{eq:EntropyDensity} and \eqref{eq:StressTensor}. 
 The most general null vector (up to a scaling) is 
\be
k^a=\left\{ 1, \frac{e_1}{a_1}, \dots, {e_d \over a_d}\right\}
\label{nullvec}
\ee
where 
\be
\sum_{i=1}^d e_i^2=1
\ee

Then
\be
s_ak^a=-s, ~~~T_{ab}k^ak^b=2\rho
\ee
Thus the condition (\ref{wcond1}) becomes
\be
s\le \sqrt{2\alpha_1\rho\over  G}
\ee
We see that we can satisfy this if we take
\be
K\le \sqrt{2\alpha_1}
\ee

\bigskip

(ii)  We have
\bea
&&s_{t;t}=-\dot s, ~~~s_{t;j}=s_{j;t}=0\nn
 &&s_{j;i}= \delta_{ij}{a_j \dot a_j} s = \delta_{ij}{C_j\over t} a_j^2\,  s
\eea
Thus the condition (\ref{wcond2}) is
\be
\left|-\dot s+\sum_{j=1}^d{e_j^2 C_j\over t} s\right|\le 2\alpha_2\rho
\label{wcond3}
\ee
We have from (\ref{qqtwo})
\bea
s&=&K\sqrt{\rho\over G}={K\over G t}{(1-\sum_i C_i^2)^\h\over \sqrt{16\pi}}, \nonumber \\
\dot s&=& -{K\over G t^2}{(1-\sum_i C_i^2)^\h\over \sqrt{16\pi}}
\eea
Then  \eqref{wcond3} becomes equivalent to the requirement
\be \label{eq:Condition1}
r''\equiv {2K\sqrt{\pi}(1+\sum_{j=1}^d e_j^2 C_j)\over \alpha_2 (1-\sum_{i=1}^d C_i^2)^\h}\le 1
\ee
We note that
\be
{r''\over r}={2\pi(1-C_1)(1+\sum_{j=1}^d e_j^2 C_j)\over \alpha_2 (1-\sum_{i=1}^d C_i^2)}
\ee
where $r$ is given by (\ref{req}). From (\ref{relationa}) we note that $\alpha_2\le \pi$. We then have
\bea
{r''\over r}&\ge &{2(1-C_1)(1+\sum_{j=1}^d e_j^2 C_j)\over  (1-\sum_{i=1}^d C_i^2)}\nn
&\ge &{2(1-C_1)(1+\sum_{j=1}^d e_j^2 C_j)\over  (1-C_1^2)}\nn
&=&{2\(1+\sum_{j=1}^d e_j^2 C_j\)\over (1+C_1)}
\eea
We can choose out null vector $k^a$ such that $e_1=1,\,  e_j=0, \, j=2, \dots d$. Then we get
\be
{r''\over r}\ge 2
\ee
Thus we see that whenever we choose the initial conditions in our cosmology to get $r>1$, we also find that $r''>1$, in violation of the condition (\ref{wcond2}) which was assumed in \cite{fmw}. Thus the situations where we violate the covariant entropy bound are not covered by the proof of \cite{fmw}.

\subsection{Other proofs}

In \cite{bfm2} another proof of the bound was given, under the assumption that the entropy current $s^a$ vanishes on the surface ${\cal S}$. But in our cosmology, $s^a$ is nonvanishing on ${\cal S}$, so this proof does not apply. 

 For situations with weak gravity, a proof of the covariant entropy bound was given in \cite{bm1,bm2}. This  proof builds on   definitions of the entropy $S$ and the energy $E$ formulated in terms of entanglement entropy \cite{casini}. It may be that the definitions of energy and entropy that give the covariant entropy bound are not the ones that are directly related to the entropy and energy that we have used. We have treated these quantities as continuous functions, defined by an appropriate local averaging proceedure. It may be that in certain situations the quantum fluctuations are such that they do not permit such an averaging, and then one may be able to preserve the covariant entropy bound in some way. But in that case it is not clear what the significance of the bound is. One would hope to use the bound in situations like the early Universe, and if the quantum fluctuations here are such that they do not give well defined local entropy and energy densities, then it is unclear how the bound could be useful.\footnote{We thank R. Brustein for a discussion on this issue.}
  
\section{The causal entropy bound}\label{ceb}

In this section we address the `causal entropy bound' proposed in \cite{brusv}. This bound is based on the idea that there is a critical `Jeans length' scale above which perturbations in a cosmology would be causally disconnected, and so  black holes larger than this scale would not form. The bound is then obtained from the entropy of the largest hole that can in fact form. The bound is expressed in a covariant form, described as follows. 

Consider  a hypersurface $\tau=constant$ in spacetime, and a region described as $\sigma<0$ on this hypersurface. Then the entropy $S$  in this region is bounded by
\bea
&&S_{CEB}\nn
&=&{c_1\over 4G}\int_{\sigma<0} d^{4} x \sqrt{-g} \delta (\tau)\sqrt{{\rm Max}_\pm \left [ (G_{\mu\nu}\pm R_{\mu\nu}) \p^\mu \tau \p^\nu \tau\right ] }\nn
&=&{c_1\sqrt{\pi\over 2G}}\int_{\sigma<0} d^{4} x \sqrt{-g} \delta (\tau)\nn
&&~~~~~~~\times\sqrt{{\rm Max}_\pm \left [ (T_{\mu\nu}\pm T_{\mu\nu}\mp \h g_{\mu\nu} T ) \p^\mu \tau \p^\nu \tau\right ] }\nn
\label{bvf}
\eea
We have included an extra constant $c_1$ of order unity (not present in the expression of \cite{brusv}), since the arguments leading to (\ref{bvf}) are order of magnitude estimates and it is not clear if they imply $c_1=1$; further, we would like to extend (\ref{bvf}) to all space dimensions $d$, and it may be that $c_1$ depends on $d$. 

Let us consider  this bound for our metric 
\be
ds^2=-dt^2+\sum_{i=1}^d a_i^2(t) dx_i dx_i
\label{metric4}
\ee
Setting $\tau=t$, we find that the delta function reduces to a 3-dimensional one over a spatial volume; thus the bound is given in terms of a volume rather than an area. This is in line with the arguments of the present paper, which suggest that bounds in terms of an area may not be valid. We have
\be
\p^\mu \tau= \{ -1, 0, 0, 0 \}
\ee
We then find 
\bea
&&{\rm Max}_\pm\left [ (T_{\mu\nu}\pm T_{\mu\nu}\mp \h g_{\mu\nu} T ) \p^\mu \tau \p^\nu \tau\right ] \nn
&=&{\rm Max}_\pm \left [ (\rho \pm \rho \pm \h  (-\rho+3\, p ) \right ] \nn
&=&{\rm Max}_\pm \left [ (\rho \pm \rho \pm \rho \right ] \nn
&=& 3 \rho
\eea
where in the third line we have set $p=\rho$, and in the last line we see that the upper signs give the required maximum. We then find
\be
S_{CEB}=3c_1\sqrt{\pi\over 2}\sqrt{\rho\over G}
\ee
The equation of state (\ref{density}) is not in contradiction with this bound, as long as 
\be
K\le 3c_1\sqrt{\pi\over 2}
\ee
If we conjecture that the causal entropy bound is saturated in the early universe by the `black hole gas' type of states, then the equation of state for this phase would be 
\be
s=3c_1\sqrt{\pi\over 2}\sqrt{\rho\over G}
\ee

\section{Discussion}\label{discussion}

We have seen that the equation of state (\ref{density}) leads to a violation of the covariant entropy bound if we take a sufficiently asymmetric cosmology. Let us now discuss the possible physical implications of this observation.

\subsection{Possible inadmissibility of the set-up}\label{discussiona}

We have made several implicit assumptions in our treatment of the equation of state; let us now see if these could be invalid:

\bigskip

(a) We required the cosmology to have an anisotropic expansion, with an   anisotropy of order unity. It is possible that the brane states depicted in fig.\ref{ffour2}(b) are resistant to such an anisotropic expansion. Consider the setup of section\,\ref{infinitesimal}. Suppose there was a dynamical reason which enforced the requirement

\be
{b_{10}\over b} ~\stackrel{??}{\le} ~{[(d-1)-{K^2\over \pi} (d-2)]\over {2K^2\over \pi}} 
\label{bbn}
\ee
Then we cannot have the asymmetry required in (\ref{bb}), and we would not be able to violate  the covariant entropy bound.  We do not however know of a clear reason that would yield a requirement like (\ref{bbn}).

\bigskip

(b) We have seen that the entropy (\ref{oneqq}) can be realized by a gas of black holes \cite{bf,veneziano}, and that in string theory we can get the same entropy by sets of intersecting branes \cite{alimathur}. It is possible that some long distance effect  disallows such an array of black holes or intersecting brane sets. In that case we may get less entropy than that given by (\ref{oneqq}), and thus not end up violating the bound. 

There are some difficulties in such a resolution however. For one thing, it is not clear what effect in general relativity could prevent us from taking a lattice of black holes. If we use intersecting brane sets in string theory, then in the configuration of fig.\ref{ffour2}(a), these brane sets are expected to just reproduce the entropy of black holes. Away from the black hole horizon the physics of these branes is supposed to approximate the usual supergravity behavior of black holes. So again it is not clear how we could avoid getting configurations that yield an entropy (\ref{oneqq}). 

In spite of these issues, let us assume that there is {\it some} long distance effect that makes the entropy of an intersecting brane lattice  less than (\ref{oneqq}). Note that it would not help if this long distance effect just reduced the value of $K$; we have seen that the bound can be violated for {\it any} value of $K$, however small. We {\it can} avoid violating the bound if (\ref{oneqq}) was modified to
\be
S=K\sqrt{EV\over G}\, \(\log {EV\over G}\)^{-1}
\label{mod}
\ee
In \cite{alimathur} it was noted when arriving at (\ref{oneqq}) that we are working in a regime ${EV\over G} \gg 1$. Thus the $\log$ factor in this relation does not vanish and lead to a divergence.  For macroscopic volumes $V$ the log reduces $S$ by a large factor. To compare this modified expression of $S$ with the expression (\ref{oneqq}), we can say that the prefactor $K$ in (\ref{oneqq}) gets replaced by an effective prefactor  $K/ \log(EV/G)$. Then eq.(\ref{c1bound}) tells us that  the values of $C_1$ needed to violate the bound tend to unity. From (\ref{qqtwo}) we then find that $\rho\rightarrow 0$, and we do not get a sensible evolution. Thus we avoid violating the covariant entropy bound.  

But a modification of the form (\ref{mod}) does not respect the underlying idea used in \cite{alimathur}  to conjecture the equation of state (\ref{oneqq}). Recall the requirement (iii) in the approach (c) mentioned in section \ref{onea}. This requirement says that the entropy $S$ in the box should be of order the Bekenstein entropy $S_{\rm bh}$ when the energy $E$ in the box is order $E_{\rm bh}$, the energy of a hole with radius the order of the box size. Since a round hole cannot exactly fit in our toroidal box, this requirement $S\sim S_{\rm bh}$ can only be an order of magnitude relation, undefined upto a factor of order unity. But this lack of precision cannot be extended to accommodate a modified relation of the form 
\be
S\sim S_{\rm bh}/\log (EV/G)\ll S_{\rm bh}
\ee
since it {\it is} in fact possible to make  a black hole in the box with $S\sim S_{\rm bh}$. Thus (\ref{mod}) does not appear to be a reasonable modification of (\ref{oneqq}); other attempts to reduce $S$ will face a similar difficulty.

\subsection{Possible constraints on applicability of the bound}

It might be that a naive use of the equation of state (\ref{density}) violates the covariant entropy bound, but that there are subtle conditions on when we can use the bound, and these prevent us from arguing that the bound is indeed violated. It has already been noted that the quantum evaporation of black holes gives a stress tensor that  requires modification of the Bousso bound \cite{lowe,strominger}.  Let us examine some possibilities related to our current matter model:

\bigskip

(a) When we model our matter as a `black hole gas' (following \cite{bf}), then these holes have some typical radius $R_s$ that depends on the elapsed time $t$ since  the big bang. A similar `lumpiness' on the scale $R_s$ may exist when the holes are replaced by sets of intersecting branes. One finds that \cite{bf,alimathur}
\be
R_s\sim t
\ee
so the size of the `matter bits' is of order the extent of the light sheet. It is possible that in this situation we cannot use the bound; the bound may require the matter fluid to be very homogenous on the scale of the light sheet.\footnote{We thank Erick Weinberg for a discussion on this point.}

If such is the case, then we learn something interesting about the domain of validity of the bound. It is also not clear if we can  use such a  bound  in the early universe, where  models like  the `black hole gas'  may describe the matter content. 

\bigskip

(b) Suppose the matter is described by the black hole gas. If we follow a null geodesic back from our hypersurface ${\cal S}$, then this geodesic can fall into one of the holes and end  up at the central singularity of the hole. In that case one may say that we cannot follow the geodesic all the way back to the cosmological singularity at $t=0$, and thus we cannot get the full light sheet that has been used above. Then we would not get all   the entropy $S_{\rm sheet}$ in (\ref{ssheet}) to flow through the sheet, and we may not be able to violate the bound. 

But in string theory we would replace the black holes by intersecting brane sets, which give rise to `fuzzballs' but no horizons or singularities. One may still argue that null geodesics follow very complicated paths when they enter an entity like a fuzzball, and one might find caustics before reaching  $t=0$. If we use such an argument to save the bound, however, then we have to ask how we can ever hope to use the bound for cosmology, since the matter in the very early universe is likely to be dense and stringy anyway.

\subsection{Replacing the area bound by a volume bound}\label{discussionc}

Finally one must consider a fundamental change in how we expect entropy bounds to work. Black holes have an entropy given by their surface area: 
\be
S={A\over 4G}
\label{entropyq}
\ee This fact has suggested that we try to bound the entropy in a region ${\cal R}$  in terms of the area of some surface associated to ${\cal R}$; the covariant entropy bound is one such conjecture. But duality symmetries of string theory suggest the answer (\ref{oneqq}), which gives the entropy in terms of the {\it volume} of ${\cal R}$. In fact, as we have noted
\be
S=K\sqrt{EV\over G}=K\sqrt{\rho\over G} \, V
\label{volumeq}
\ee
so the entropy is extensive in the volume just as for normal thermodynamic systems. The area entropy (\ref{entropyq}) arises as a particular limit, where the energy $E$ is just enough to make a single black hole in the box of volume $V$.  This can be seen as follows. Suppose we have just enough energy to make a single black hole radius the size of the box:   
\be
R_s\sim V^{1\over d}
\ee
The energy of this hole is (eq.(\ref{eseven}))
\be
E\sim {R_s^{d-2}\over G}\sim {V^{d-2\over d}\over G}
\ee
The entropy expression $S\sim \sqrt{EV\over G}$ gives
\be
S\sim \sqrt{V^{{d-2\over d}+1}\over G^2}\sim {V^{d-1\over d}\over G}\sim {A\over G}
\label{areaww}
\ee
Thus we recover the area dependence of (\ref{entropyq}). If we have {\it more} energy in the box, then we will have several holes, and the expression $\sim \sqrt{EV\over G}$ will cover this more general case as well. As noted in \cite{alimathur}, we expect the volume dependent expression (\ref{volumeq}) to hold over the domain
\be
\rho_{\rm bh}\lesssim \rho\lesssim \rho_p
\label{twoq}
\ee
Here $\rho_{\rm bh}$ is the energy density that is just enough to make a single black hole of order the box size; the entropy at this density matches onto the area expression as noted in (\ref{areaww}).  At the upper end $\rho=\rho_p$  (\ref{volumeq}) gives an entropy of one bit per unit Planck volume. This corresponds to having a lattice of black holes with each hole of order Planck size. Thus (\ref{volumeq}) extrapolates the Bekenstein `area entropy'  \cite{bek} to the domain (\ref{two}). Since $\rho \gtrsim \rho_{\rm bh}$, we will say that matter is `hyper-compressed'; i.e., compressed beyond the density of the largest black hole that can fit in the box.

We have seen that the causal connecton bound of \cite{brusv} expresses the entropy in a region in terms of the volume of region, rather than area of its boundary. We have seen that our arguments do not rule out this bound; rather they support the spirit in which this bound was proposed.

It is interesting that the Bekenstein bound \cite{bek2} related the entropy to the energy $E$ and the parameter $R$ giving the {\it linear} size of the system. The covariant entropy bound tries to extend the behavior of black hole entropy to general situations, limiting the entropy through a light sheet in terms of the {\it area} $A$ of the enclosed region. The expression (\ref{volumeq}) gives an entropy that is proportional to the {\it volume} $V$.\footnote{In \cite{vgu} it was argued that the expression for the entropy of a black hole should be modified in a cosmological context, to one which involves length and volume terms besides the area.}

 It would be interesting to explore these general ideas further,  in particular using  lessons from the fuzzball picture of black holes. This picture suggests that  whenever there is enough mass in a region to create a horizon, then tunneling into fuzzballs leads to a breakup of the spacetime into parts that are disconnected but entangled \cite{breakup}. This suggests that the black hole gas  may have to thought of as a collection of disjoint spacetime regions, that join up as the expansion proceeds. We hope to return to these considerations elsewhere.\footnote{For other conjectures on the evolution of the big bang/ big crunch, see \cite{brus2000}.} 

\appendix 
\section{General power-law entropy expression}
\label{Appendix1}
In this appendix we study a general power-law entropy expression in an isotropic  spatially flat FRW universe. The entropy for a photon gas or the entropy given by  Eq.\eqref{density}  would be special cases of this general expression.  We work in ($d$+1) spacetime dimensions. We assume the the entropy density has the form
\begin{equation} \label{eq:GeneralS}
	s= m^\alpha \rho^{\mu }~.
\end{equation}
Here $m$ is a constant  with the units of mass.  (In most cases $m$ would just be the   Planck mass). Matching dimensions, we find
\begin{equation}\label{eq:RightUnits}
	\alpha = d(1-\mu)-\mu~.
\end{equation}
We first derive the thermodynamical properties of this equation of state: 
\begin{eqnarray} \label{eq:ThermoRels}
	T&=& \left(\frac{\partial U}{\partial S}\right)_V= \left(\frac{\partial (U/V)}{\partial (S/V)}\right)_V=\left(\frac{\partial\rho}{\partial s}\right)_V= \frac1{\mu m^\alpha} \rho^{1-\mu}~,\cr \cr
	p&=& T\left(\frac{\partial S}{\partial V}\right)_E= T(1-\mu)m^\alpha \rho^\mu= \frac{1-\mu}{\mu} \rho~.
\end{eqnarray}
Thus
\begin{equation}\label{eq:GeneralW}
	w=\frac{(1-\mu)}\mu~,
\end{equation}
As expected, this  gives $w=1$ for $\mu=\frac12$. In the Friedmann equations  \eqref{gtt2} and \eqref{gkkpre2}, we assume for simplicity that we have isotropy:
  \be
  a_i(t)=a(t), ~~~~~~i=1, \dots d
  \ee  
  Then we get 
\begin{equation}\label{eq:GeneralFriedmann}
	\frac {a \ddot a}{\dot a^2}=- \frac{w d + d -2}{2}= 1- \frac{d}{2\mu}~.
\end{equation}
The scale factor and entropy density are given by 
\begin{eqnarray} \label{eq:GeneralPowerLawSol}
	&&a(t) = a_0 t^{2\mu / d}~, \qquad \rho = \frac{\mu^2 (d-1)}{4 \pi G  t^2 d}~,\cr \cr
	&&s(t) = m^\alpha \left( {\mu^2 (d-1)\over 4 \pi G d }\right)^\mu t^{-2 \mu} \equiv \eta t^{-2\mu}~.
\end{eqnarray}

Now consider the computation of entropy passing through a lightsheet, as carried out in section\,\ref{secchecka}. 
As in that computation, we  choose the surface $\cal S$ at a time $t_0$, in a plane orthogonal to the $x^1$ direction. Using  the results from \eqref{eq:PhotonVolumeElem} and repeating the steps in the computation leading to \eqref{ssheet} gives
\begin{eqnarray}\label{eq:PhotonVolumeElem}
	S_{\rm sheet }&=& \int_0^{t_0}s(t) { A} \prod_{i=2}^d\left( {a_i(t)\over a_i(t_0)}\right) a_1(t){dx^1 \over dt} dt\cr \cr &=& \eta  \int_0^{t_0}{t_0^{2 \mu (1-d) \over d}} t^{-{2\mu \over d}} { A}dt~. 
\end{eqnarray}
To get a finite value for the entropy passing through the sheet we need 
\begin{equation}\label{eq:Constraint}
	2\mu <d~.
\end{equation}
We notice that for $\mu <\frac{1}{2}$, the ratio of  $S_{\rm sheet}$ to the  Bousso bound increases without any limit. Our equation of state (\ref{density}) has   $\mu = \frac12$, and we see it is a marginal case in this sense. For the cases $\mu>\frac12$ we have
\begin{equation}\label{eq:entropyRatio}
	{\rm r}= \frac{S_{\rm sheet}}{S_{\rm bound}}=\frac{4 G\eta d}{  d - 2 \mu} t_0^{1-2\mu}.
\end{equation}
Thus we can formally  violate the bound by going to a sufficiently small value of $t_0$; that is by choosing
\begin{equation}\label{eq:ViloationCondition}
t_0 < {\cal F}^{1/(1-2\mu)} M_P^{(1-d)(1-\mu)/(1-2\mu)} m^{\alpha/(1-2\mu)}~,
\end{equation}
where 
\begin{eqnarray} \label{eq:FDef}
	{\cal F} &=& \frac{4 d }{(d-2\mu)} \left( \frac{\mu^2 (d-1)}{4\pi d} \right)^\mu
\end{eqnarray}
and we have set $G={1 /M_{\rm P}^{d-1}}$. 
If the scale $m$ is set at the Planck mass, then it is easy to check that the  condition (\ref{eq:ViloationCondition}) gives  $t_0\le t_P$. Since we do not expect our gravitational analysis to make sense at times earlier than Planck time, we see that there is no violation of the bound for $\mu>\h$.

\section{Entropy in a comoving volume } \label{sect:ComovingEntropy}

In this appendix we show that for the equation of state (\ref{density}), the entropy enclosed in any comoving volume is constant during the cosmological evolution. Consider  a small comoving  cubic coordinate volume $\Delta V_{\rm coord}=\Delta x_1\Delta x_2 \ldots \Delta x_d$. The entropy enclosed is $s \Delta V_{\rm phys}$ where $s$ is the entropy density given by \eqref{density} and $\Delta V_{\rm phys}$ is the physical enclosed volume: 
\begin{equation}\label{eq:TwoDeltaV's}
\Delta V_{\rm phys}(t) = \Delta V_{\rm coord}\prod_{i=1}^d a_i(t) ~.
\end{equation}
Using \eqref{condition} we get 
\begin{eqnarray} \label{eq:isoEnt}
	\Delta S_{\rm encl} (t) &=& s(t) \Delta V_{\rm phys}(t) =  \frac Kt\sqrt{\alpha \over G}\,   \Delta V_{\rm coord}\prod_{i=1}^d a_i(t) \cr \cr 
	&=& \frac Kt\sqrt{\alpha \over G} \,  \Delta V_{\rm coord}\left( \prod_{i=1}^d a_{i0}\right) t^{\sum_{i=1}^d C_i}\nonumber \\ 
	&=& K \sqrt{\alpha \over G} \,  \Delta V_{\rm coord}\left( \prod_{i=1}^d a_{i0}\right) ~.
\end{eqnarray}
where we have used (\ref{condition}). Thus the entropy in our comoving volume  is  independent of $t$.

   \bigskip
   
   \bigskip 
    
\begin{acknowledgments}
We thank  R. Bousso,  R. Brandenberger,  R. Brustein, Sumit Das, V. Jejjala, S. Kalyana Rama, Don Marolf, Rob Myers, Soo-Jong Rey, Ken Olum, G. Veneziano, Robert Wald and Erick Weinberg for  discussions.
This work was supported in part by DOE grant DE-FG02-91ER-40690 and National Science Foundation (grant PHY-1213888).
\end{acknowledgments}

\nocite{*}


\begin{thebibliography}{00}


\bibitem{bek2} 
  J.~D.~Bekenstein,
  Phys.\ Rev.\ D {\bf 23}, 287 (1981).
   
    \bibitem{bek}
J.~D.~Bekenstein,
Phys.\ Rev.\ D {\bf 7}, 2333 (1973).
%

  
\bibitem{bek3} 
  J.~D.~Bekenstein,
  Int.\ J.\ Theor.\ Phys.\  {\bf 28}, 967 (1989).
  
  
\bibitem{fs} 
  W.~Fischler and L.~Susskind,
  hep-th/9806039.
  
  
\bibitem{venezianopre} 
  G.~Veneziano,
  Phys.\ Lett.\ B {\bf 454}, 22 (1999)
  [hep-th/9902126].
  
\bibitem{veneziano2} 
  G.~Veneziano,
  hep-th/9907012.
 
\bibitem{br} 
  D.~Bak and S.~-J.~Rey,
  Class.\ Quant.\ Grav.\  {\bf 17}, L83 (2000)
  [hep-th/9902173].
  
  
\bibitem{bousso} 
  R.~Bousso,
  JHEP {\bf 9907}, 004 (1999)
  [hep-th/9905177];
 R.~Bousso,
  Rev.\ Mod.\ Phys.\  {\bf 74}, 825 (2002)
  [hep-th/0203101].
  
  
\bibitem{veneziano} 
  G.~Veneziano,
  JCAP {\bf 0403}, 004 (2004)
  [hep-th/0312182].

  
  
\bibitem{el} 
  R.~Easther and D.~A.~Lowe,
  Phys.\ Rev.\ Lett.\  {\bf 82}, 4967 (1999)
  [hep-th/9902088].
  
\bibitem{rs} 
  S.~Kalyana Rama and T.~Sarkar,
  Phys.\ Lett.\ B {\bf 450}, 55 (1999)
  [hep-th/9812043].
  
\bibitem{brusv}
  R.~Brustein and G.~Veneziano,
  Phys.\ Rev.\ Lett.\  {\bf 84} (2000) 5695
  [hep-th/9912055].

  \bibitem{sv}
A.~Strominger and C.~Vafa,
Phys.\ Lett.\ B {\bf 379}, 99 (1996)
[arXiv:hep-th/9601029];
%
  G.~T.~Horowitz, D.~A.~Lowe and J.~M.~Maldacena,
  Phys.\ Rev.\ Lett.\  {\bf 77}, 430 (1996)
  [arXiv:hep-th/9603195];
  C.~V.~Johnson, R.~R.~Khuri and R.~C.~Myers,
  Phys.\ Lett.\ B {\bf 378}, 78 (1996)
  [arXiv:hep-th/9603061].


\bibitem{hms}
  G.~T.~Horowitz, J.~M.~Maldacena and A.~Strominger,
  Phys.\ Lett.\ B {\bf 383}, 151 (1996)
  [arXiv:hep-th/9603109].

  
    \bibitem{fuzzballsize}
  O.~Lunin and S.~D.~Mathur,
  Phys.\ Rev.\ Lett.\  {\bf 88}, 211303 (2002)
  [hep-th/0202072].


 
\bibitem{bf}
  T.~Banks and W.~Fischler,
  hep-th/0102077;
    T.~Banks and W.~Fischler,
  hep-th/0212113;
T.~Banks and W.~Fischler,
  Phys.\ Scripta T {\bf 117}, 56 (2005)
  [hep-th/0310288];
 T.~Banks and W.~Fischler,
  hep-th/0405200;
  T.~Banks and W.~Fischler,
  hep-th/0412097.
  

\bibitem{hp}
  G.~T.~Horowitz and J.~Polchinski,
  Phys.\ Rev.\ D {\bf 55}, 6189 (1997)
  [arXiv:hep-th/9612146];
G.~T.~Horowitz and J.~Polchinski,
  Phys.\ Rev.\ D {\bf 57}, 2557 (1998)
  [arXiv:hep-th/9707170].


\bibitem{sas1}
  N.~Sasakura,
  Prog.\ Theor.\ Phys.\  {\bf 102}, 169 (1999)
  [hep-th/9903146].


  
\bibitem{brus2} 
  R.~Brustein, S.~Foffa and A.~E.~Mayo,
  Phys.\ Rev.\ D {\bf 65}, 024004 (2002)
  [hep-th/0108098].

\bibitem{verlinde} 
  E.~P.~Verlinde,
  hep-th/0008140.


  
\bibitem{sas2} 
  N.~Sasakura,
  Phys.\ Lett.\ B {\bf 550}, 197 (2002)
  [hep-th/0209220];






  
\bibitem{alimathur} 
  A.~Masoumi and S.~D.~Mathur,
  Phys.\ Rev.\ D {\bf 90}, no. 8, 084052 (2014)
  [arXiv:1406.5798 [hep-th]].


\bibitem{bv} 
  R.~H.~Brandenberger and C.~Vafa,
  Nucl.\ Phys.\ B {\bf 316}, 391 (1989).

\bibitem{branegases}
S.~Alexander, R.~H.~Brandenberger and D.~Easson,
  Phys.\ Rev.\ D {\bf 62}, 103509 (2000)
  [arXiv:hep-th/0005212];
  R.~Easther, B.~R.~Greene, M.~G.~Jackson and D.~Kabat,
  Phys.\ Rev.\ D {\bf 67}, 123501 (2003)
  [arXiv:hep-th/0211124].

\bibitem{rama} 
 S.~Kalyana Rama,
  Phys.\ Lett.\ B {\bf 638}, 100 (2006)
  [hep-th/0603216];
  S.~Kalyana Rama,
  Phys.\ Lett.\ B {\bf 645}, 365 (2007)
  [hep-th/0610071].


\bibitem{marolf} 
  D.~Marolf and R.~D.~Sorkin,
  Phys.\ Rev.\ D {\bf 69}, 024014 (2004)
  [hep-th/0309218];
  Phys.\ Rev.\ D {\bf 69}, 064006 (2004)
  [hep-th/0310022];
  D.~Marolf,
  hep-th/0410168.



 
\bibitem{bfm} 
  T.~Banks, W.~Fischler and L.~Mannelli,
  Phys.\ Rev.\ D {\bf 71}, 123514 (2005)
  [hep-th/0408076].

\bibitem{kl} 
  N.~Kaloper and A.~D.~Linde,
  Phys.\ Rev.\ D {\bf 60}, 103509 (1999)
  [hep-th/9904120].

\bibitem{fmw} 
  E.~E.~Flanagan, D.~Marolf and R.~M.~Wald,
  Phys.\ Rev.\ D {\bf 62}, 084035 (2000)
  [hep-th/9908070].


\bibitem{cm} 
  B.~D.~Chowdhury and S.~D.~Mathur,
  Class.\ Quant.\ Grav.\  {\bf 24}, 2689 (2007)
  [hep-th/0611330].
  
\bibitem{son} 
  P.~Kovtun, D.~T.~Son and A.~O.~Starinets,
  Phys.\ Rev.\ Lett.\  {\bf 94}, 111601 (2005)
  [hep-th/0405231].

  

  
\bibitem{bfm2} 
  R.~Bousso, E.~E.~Flanagan and D.~Marolf,
  Phys.\ Rev.\ D {\bf 68}, 064001 (2003)
  [hep-th/0305149].

\bibitem{bm1} 
  R.~Bousso, H.~Casini, Z.~Fisher and J.~Maldacena,
  arXiv:1404.5635 [hep-th];
  
  \bibitem{bm2}
  R.~Bousso, H.~Casini, Z.~Fisher and J.~Maldacena,
  arXiv:1406.4545 [hep-th].
  
  
\bibitem{casini} 
  H.~Casini,
  Class.\ Quant.\ Grav.\  {\bf 25}, 205021 (2008)
  [arXiv:0804.2182 [hep-th]].

  
  \bibitem{hawking}
  S.~W.~Hawking,
  ``Particle Creation By Black Holes,''
  Commun.\ Math.\ Phys.\  {\bf 43}, 199 (1975)
  [Erratum-ibid.\  {\bf 46}, 206 (1976)];
  S.~W.~Hawking,
  ``Breakdown Of Predictability In Gravitational Collapse,''
  Phys.\ Rev.\  D {\bf 14}, 2460 (1976).

\bibitem{lowe} 
  D.~A.~Lowe,
  JHEP {\bf 9910}, 026 (1999)
  [hep-th/9907062].
  

\bibitem{strominger} 
  A.~Strominger and D.~M.~Thompson,
  Phys.\ Rev.\ D {\bf 70}, 044007 (2004)
  [hep-th/0303067].
  
\bibitem{vgu} 
  S.~Viaggiu,
  Mod.\ Phys.\ Lett.\ A {\bf 29}, 1450091 (2014)
  [arXiv:1405.6816 [hep-th]].

  
\bibitem{breakup} 
  S.~D.~Mathur,
  arXiv:1402.6378 [hep-th].
  
\bibitem{brus2000} 
  R.~Brustein and M.~Schmidt-Sommerfeld,
  JHEP {\bf 1307}, 047 (2013)
  [arXiv:1209.5222 [hep-th]].



 \end{thebibliography}

\end{document}